\documentclass[%
preprint,
superscriptaddress,
longbibliography,
 amsmath,amssymb,
 aps,
 prb,
]{revtex4-1}
\setcitestyle{numbers,square}

\usepackage[utf8]{inputenc}
\usepackage[T1]{fontenc}
\usepackage{braket}
\usepackage{graphicx}% Include figure files
\usepackage{dcolumn}% Align table columns on decimal point
\usepackage{bm}% bold math
\usepackage[normalem]{ulem} % strikethrough
\usepackage{xcolor}
\usepackage{hyperref}% add hypertext capabilities
\usepackage{siunitx}
\usepackage{mathtools}

\usepackage{pgffor}

\hypersetup{
    colorlinks,%
    citecolor=blue,%
    linkcolor=blue,%
    urlcolor=blue
}

\newcommand{\orcid}[1]{\href{https://orcid.org/#1}{\includegraphics[width=8pt]{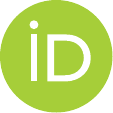}}}
\begin{document}

\title{Heterogeneous Molecular Signatures of Human Odor Perception}

\author{P. Zanineli \orcid{0009-0008-2359-5218}}
\affiliation{Brazilian Nanotechnology National Laboratory (LNNano), Brazilian Center for Research in Energy and Materials (CNPEM), Campinas, SP, Brazil.}
\affiliation{Center for Natural and Human Sciences, Federal University of ABC, Santo André, SP 09280-560, Brazil}

\author{E. V. C. Lopes \orcid{0000-0002-7981-2161}}
\affiliation{Ilum School of Science, Brazilian Center for Research in Energy and Materials (CNPEM), Campinas, SP, Brazil.}

\author{G. R. Schleder \orcid{0000-0003-3129-8682}}
\affiliation{Brazilian Nanotechnology National Laboratory (LNNano), Brazilian Center for Research in Energy and Materials (CNPEM), Campinas, SP, Brazil.}
\affiliation{Center for Natural and Human Sciences, Federal University of ABC, Santo André, SP 09280-560, Brazil} 

\author{L. N. Lemos \orcid{0000-0002-0898-568X}}
\affiliation{Ilum School of Science, Brazilian Center for Research in Energy and Materials (CNPEM), Campinas, SP, Brazil.} 

\author{F. Crasto de Lima \orcid{0000-0002-2937-2620}} 
\email{felipe.lima@ilum.cnpem.br}
\affiliation{Ilum School of Science, Brazilian Center for Research in Energy and Materials (CNPEM), Campinas, SP, Brazil.}

\author{A. Fazzio \orcid{0000-0001-5384-7676}}
\email{adalberto.fazzio@ilum.cnpem.br}
\affiliation{Ilum School of Science, Brazilian Center for Research in Energy and Materials (CNPEM), Campinas, SP, Brazil.}
\affiliation{Center for Natural and Human Sciences, Federal University of ABC, Santo André, SP 09280-560, Brazil}

%%%%%%%%%%%%%%%%%%%%%%%%%%%%%%%%%%%%%%%%%%   ABSTRACT 

\begin{abstract}

Understanding how molecular structure gives rise to odor perception remains a long-standing challenge, with ongoing debate over whether olfaction is primarily governed by molecular shape, vibrational properties, or their interplay at the level of olfactory receptors.
Here, we ask whether different odors rely on common molecular determinants or instead emerge from distinct physicochemical regimes.
Using interpretable machine-learning models trained on molecular descriptors derived from first-principles calculations that span electronic, vibrational, and structural properties, we analyze feature contributions for odor categories and their associated receptors.
We find that no single descriptor class universally dominates odor prediction; instead, different odors exhibit strongly odor-specific patterns of feature importance, with substantial variability across physicochemical domains.
This heterogeneity is consistent across different models, suggesting that a universal encoding scheme does not capture odor perception but reflects receptor- and odor-dependent structure–odor relationships.
Our results provide statistical constraints on competing olfactory theories and offer a data-driven framework for organizing odor space.

\end{abstract} 

\maketitle

\section{Introduction}

The molecular basis of olfaction remains one of the most intriguing and debated topics in sensory science.
At the molecular level, this process is initiated by olfactory receptors, a large family of G protein–coupled receptors embedded in the membranes of olfactory sensory neurons, which directly interact with odorant molecules and convert their physicochemical properties into intracellular signaling events \cite{su2009olfactory}.
While the downstream transduction cascade triggered by receptor activation has been well characterized \cite{CELLbuck1991, CELLmalnic1999}, the upstream question of how molecular features are encoded at the receptor level and translated into perceptual categories remains actively investigated \cite{CSturin1996, PRLbrookes2007, billesbolle2023structural, de2024engineered, wang2026structural, han2026mechanistic}.

Several conceptual frameworks have been proposed to account for this process, ranging from classical structural complementarity models to vibrational theories of olfaction and combinatorial coding schemes \cite{CSturin1996, PRLbrookes2007}.
Recent structural and computational studies have begun to clarify important aspects of odorant recognition, suggesting that many olfactory receptors operate through flexible binding pockets, sparse polar anchoring, and substantial conformational heterogeneity of both ligand and receptor \cite{billesbolle2023structural, de2024engineered}.
Consistent with this view, cryo-EM structures of class II olfactory receptors bound to highly hydrophobic ligands reveal large, permissive binding cavities and unconventional activation pathways, supporting a model in which odor discrimination emerges from dynamic receptor–ligand interaction landscapes rather than rigid lock-and-key recognition.
At the same time, recent structural evidence indicates that this conformational framework does not exhaust the physicochemical space of olfactory recognition: in a human olfactory receptor selective for aldehydes, odorant binding involves a reversible covalent linkage that transiently integrates the ligand into the receptor’s electronic environment.
While this mechanism does not establish a quantum mode of olfaction, it provides concrete molecular grounds for considering how electronic reactivity and vibrational properties of odorants could, in principle, modulate receptor activation under specific microenvironmental conditions \cite{PRLbrookes2007, PNASgray2005, PCCPsolovyov2012, JCPchciska2015, PREtirandaz2015, Liu2020}.
How these factors interact with molecular structure and receptor dynamics, however, remains unresolved \cite{PMJbarwich2015}.

Understanding these molecular mechanisms is a technological challenge with profound implications.
The development of artificial olfactory systems, commonly referred to as electronic noses, and bio-inspired chemical sensors depends critically on elucidating how natural receptors encode and discriminate chemical information \cite{NARollitrault2024, TFSTzhong2024}.
A deeper understanding of odor–receptor interactions may guide the rational design of such devices and provide insights into the evolution of chemosensory systems, from simple microbial sensors to the complex olfactory systems of mammals \cite{SCIENCElee2023}.

Recent data-driven approaches have sought to predict olfactory perception by correlating molecular descriptors with perceptual classes of odor \cite{SCIENCElee2023, SCIENCEkeller2017, SRsaini2022, JCIMburns2023}.
Most of these studies focus on structural or topological features of the molecules, occasionally extending to global physicochemical descriptors.
However, the inclusion of electronic and vibrational properties has remained relatively limited, despite growing evidence that these features might encode essential cues for receptor activation and perceptual discrimination \cite{PRLbrookes2007, JCPchciska2015, ACRnarth2015}.
The growing availability of datasets such as QuantumScents \cite{JCIMburns2023} and MORE-Q \cite{SDchen2025} now enables more systematic investigations combining quantum-mechanical and structural information.

In this work, we present an integrated analysis that combines molecular structural, electronic, and vibrational properties to examine potential mechanisms of olfactory recognition.
We bring together three complementary datasets: (i) perceptual odor classifications for approximately 3,500 molecules from the Leffingwell PMP database \cite{Leffingwell}; (ii) a set of quantum-mechanical descriptors, including electronic, vibrational, and structural features, computed here from first-principles methods for the same molecular set; and (iii) the M2OR database \cite{lalis2024m2or}, which links odorant molecules to experimentally characterized and predicted olfactory receptor interactions.
By jointly analyzing these data, we explore how physicochemical molecular features relate to receptor activation patterns and perceived odor categories.
This integrative approach allows us to revisit the long-standing discussion of shape- and vibration-based contributions to olfaction from a quantitative, data-driven perspective, and to identify molecular attributes most strongly associated with odor recognition. More broadly, our results provide a framework to guide future experimental tests of olfactory mechanisms and to inform the development of predictive and artificial olfactory systems.

\section{Methods}

The molecule–odor relation was extracted from the Leffingwell PMP 2001 database, which contains $3,523$ molecules with expert-labeled 114 odor descriptors \cite{Leffingwell}, shown in Table~\ref{tab:odors}. From the isomeric SMILES representation of each molecule, an initial three-dimensional molecular structure was generated and subsequently optimized.

The atomic positions were fully relaxed, and the electronic and vibrational properties were computed within the framework of density functional theory (DFT) using the ORCA package \cite{JCPneese2020}. All calculations employed the hybrid B3LYP exchange–correlation functional \cite{JCPbecke1993, PRBlee1988} in combination with def2-SVP Gaussian-type orbital basis set \cite{PCCPweigend2005}. Geometry optimizations were performed under tight convergence criteria (energy change below $10^{-8}$\,Eh, maximum gradient below $3\times10^{-4}$\,Eh/bohr). Vibrational frequencies were computed numerically from finite differences of analytical gradients using the harmonic approximation as implemented in ORCA. Out of the $3,523$ molecules, $3,445$ yielded valid geometries and achieved full self-consistent-field and optimization convergence, and were retained for further analysis.

From the DFT calculations, we obtained the optimized molecular geometries. Using the RDKit package~\cite{rdkit}, we extracted a set of structural descriptors, summarized in Table~\ref{tab:feat}. In addition, the electronic structure calculations provided features related to the charge distribution, including the molecular dipole moment and the total partial charge on carbon ($C_c$), hydrogen ($H_c$), and heteroatoms ($X_c$). The frontier orbital energies (HOMO and LUMO) were also extracted and referenced to the vacuum level. For the vibrational information, the computed frequencies were grouped into 50 bins spanning the energy range from $0.0$ to $0.5$\,eV, resulting in a histogram representation denoted as $v_1, v_2, \ldots, v_{50}$. These structural, vibrational, and electronic features (total of 80 features) were used as molecular descriptors for subsequent analysis.

\begin{table*}[th!]
\begin{ruledtabular}
\caption{\label{tab:feat}List of molecular features extracted from DFT and RDKit analyses.}
\renewcommand{\arraystretch}{1.2}
\begin{tabular}{ll}

\textbf{Category} & \textbf{Features} \\
\hline
Structural &
rot1, rot2, rot3 (rotational constants), MolecularWeight, Aliphatic carbocycle fraction, \\ & Aliphatic heterocycles count, Aliphatic rings count, Amide bonds count, \\ & Aromatic carbocycles count, Aromatic heterocycles count,\\
& Stereocenters count, Bridgehead atoms count, Hydrogen bond acceptors count, \\ & Hydrogen bond donors count, Heteroatoms count, Heterocycles count, \\ & Rotatable bonds count, Saturated carbocycles count, Saturated heterocycles count, \\ & Spiro atoms count, Kier Phi value \\[4pt]

Vibrational &
\(v_1, v_2, \ldots, v_{50}\) (50-bin histogram of frequencies from 0.0 to 0.5~eV) \\[4pt]

Electronic &
\(p_1, p_2, p_3, p_t\) (dipole components and magnitude), HOMO, LUMO, \(C_c\), \(H_c\), \(X_c\) \\

\end{tabular}
\end{ruledtabular}
\end{table*}

Henceforth, we will refer to the structural, vibrational and electronic features as $s_i$, $v_i$ and $e_i$, respectively, with $i=1,2,3,\cdots$, following the order defined in Table~\ref{tab:feat}.

The Random Forest (RF) algorithm using molecular descriptors as input features to predict each odor label.
The dataset is inherently imbalanced: while some odors are associated with thousands of molecules, others occur in only a few tens of samples.
To address this issue, we applied an undersampling strategy by randomly selecting an equal number of negative samples (molecules not labeled with the odor under analysis) for each odor, thereby balancing the dataset.
A separate RF model was trained for every odor label.
The final RF models were trained using $200$ trees, a maximum depth of 10, and a minimum of five samples per leaf node to promote generalization rather than memorization.
Each split consider only the square root of the total number of descriptors.
The model evaluation was performed using 5-fold cross-validation with an 80/20 train–test split.
To assess the robustness of the predictions with respect to random sampling in the balancing procedure, this entire process was repeated $100$ times for each odor, varying the random selection of samples in each iteration.

Predicted odor-molecule associations were integrated with the M2OR database \cite{lalis2024m2or} to enable comparison with previously reported olfactory receptor–ligand relationships. Molecules and odor descriptors were matched across datasets using standardized chemical identifiers and curated nomenclature, as described previously. This integration allowed systematic cross-referencing of predicted associations with known receptor–odor mappings.

The analysis of olfactory receptor–ligand was restricted to human olfactory receptors by selecting entries annotated as \textit{Homo sapiens}. Molecular correspondence between datasets was established using two complementary matching criteria: (i) exact matching of odorant names and (ii) structure-based matching using SMILES representations. Finally, from the unified set of odorant–receptor associations, a high-confidence subset of interactions was defined by retaining only records that satisfied the following conditions: a positive receptor response (Responsive = 1), stimulation by a single odorant compound (Mixture = mono), and experimental annotation classified as secondary screening. This filtering step yielded a reduced dataset comprising odorant–receptor associations supported by consistent experimental evidence.

\section{Results}

To investigate how molecular properties contribute to odor perception, we analyzed a dataset of $3,445$ molecules annotated with 113 odor categories.
For each molecule, physically motivated descriptors spanning electronic, vibrational, and structural properties were generated from first-principles calculations and used as features in interpretable machine-learning models.
To facilitate comparison across odors, descriptors were grouped into three physicochemically defined domains, which serve as the basis for the analyses presented below.

An analysis over the correlation distance ($d$) between different odors classified in the Leffingwell database do not show a significant proximity ($d>0.4$).
That is, in average there is no correlation between mutual occurrences of such odors and they can be considered non-equivalent.
However, we can see if a giving odor correlates with the molecular features (electronic, vibrational or structural).
Computing the Pearson correlation between the molecular features and the odor we can look towards the existence of linear relations.

\subsection{Relationships and Correlations Between Odors and Molecular Features}

\begin{figure*}[th!]
    \centering
    \includegraphics[width=0.8\linewidth]{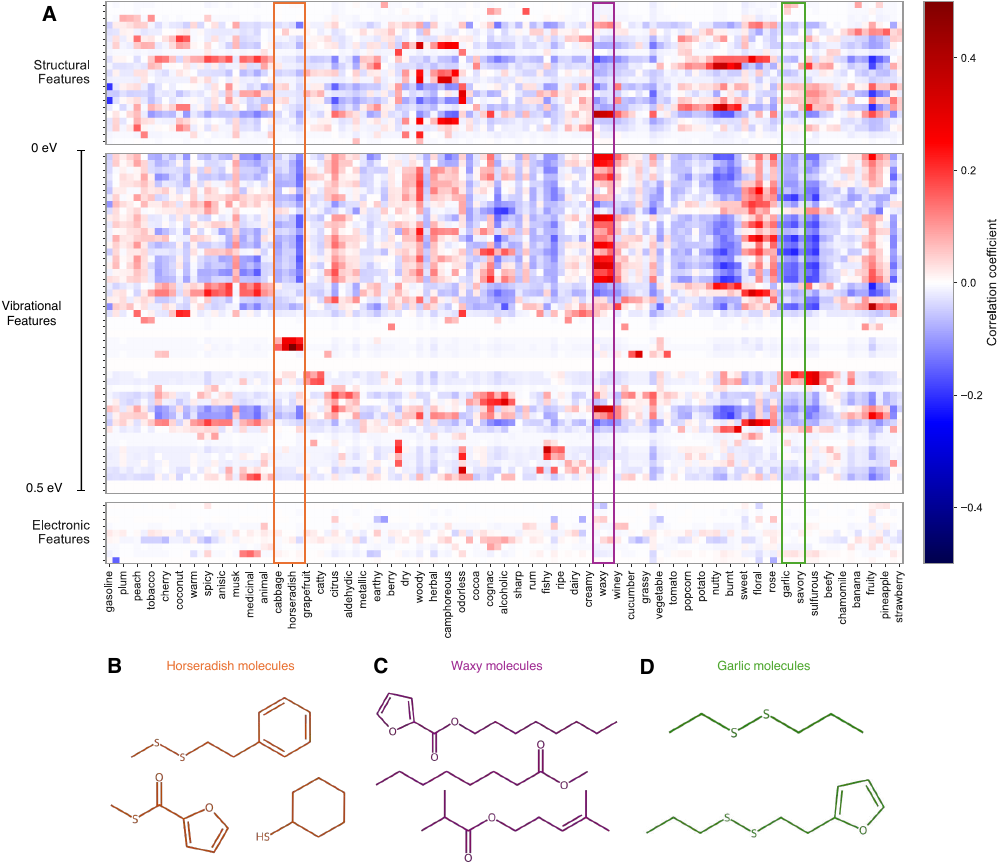}
    \caption{\label{fig:correlation}
    ({\bf A}) Heatmap showing Pearson correlations between molecular features and odor descriptors.
    The odors were ordered according their correlation distance exclusively considering their co-occurrence (see Supplemental Section S1).
    Red and blue indicate positive and negative correlations, respectively, revealing structure–odor relationships across electronic (e$j$), vibrational (v$j$), and structural (s$j$) domains.
    The color arrows highlight different molecular arrangements that are related with distinct odors. ({\bf B}), ({\bf C}), and ({\bf D}) show molecular examples of odors highlighted in ({\bf A}). 
    }
\end{figure*}

From the molecular features extracted from the  density functional theory (DFT) calculations (see Methods section), separated in electronic, vibrational and structural classes,  we compute the Pearson correlation with the odor labels.
Distinct correlation blocks highlight that related odors share similar physicochemical signatures (Fig.~\ref{fig:correlation}A).
For instance, the set of odor [cabbage, radish, horseradish, pungent] highlighted by the green section, have a positive correlation with the presence of vibrational frequencies with energy between $0.28$ and $0.3$\,eV (features v28 and v29), with example molecules in Fig.~\ref{fig:correlation}B.
Differently, the set [garlic, onion, savory], shown in the gold section, have positive correlation also with the structural features s13, s14 and s15 (respectively, Hydrogen bond acceptors, Hydrogen bond donors and Heteroatoms count), examples of molecules in Fig.~\ref{fig:correlation}D.
Among the odor categories analyzed, the set [fatty, waxy, oily] exhibited the strongest correlations.
Here positive correlation in the vibrational frequencies in the range of $0.01$ to $0.2$\,eV, but also with structural features reflecting hydrophobicity and molecular flexibility, including the aliphatic carbocycle fraction, the count of rotatable bonds, and the Kier Phi value, highlighted in purple section.
These properties characterize long, nonpolar aliphatic chains with low dipole moments and few heteroatoms --- typical of higher alcohols, fatty esters, and other compounds perceived as waxy or oily, Fig.~\ref{fig:correlation}C.
The observed patterns support a multidimensional view of olfactory coding, in which perceptual similarity arises from overlapping subspaces rather than isolated descriptors.
To capture such interactions beyond linear correlations, we next turn to non-linear machine-learning models.

%\FloatBarrier
%\FloatBarrier

\subsection{Random Forest Model Performance and Feature Importance}

To probe non-linear relationships beyond Pearson correlations, we trained Random Forest (RF) classifiers using molecular descriptors to predict each odor label independently.
Because the dataset is strongly imbalanced --- some odors occur in thousands of molecules whereas others appear in only a few tens (see Fig.~\ref{fig:s-num-data-odor}) --- we employed an undersampling strategy in which, for each odor, an equal number of negative samples was randomly selected to balance the classes.
This procedure was repeated multiple times to assess robustness against sampling variability (see Methods).

\begin{figure*}[th!]
    \centering
    \includegraphics[width=0.8\linewidth]{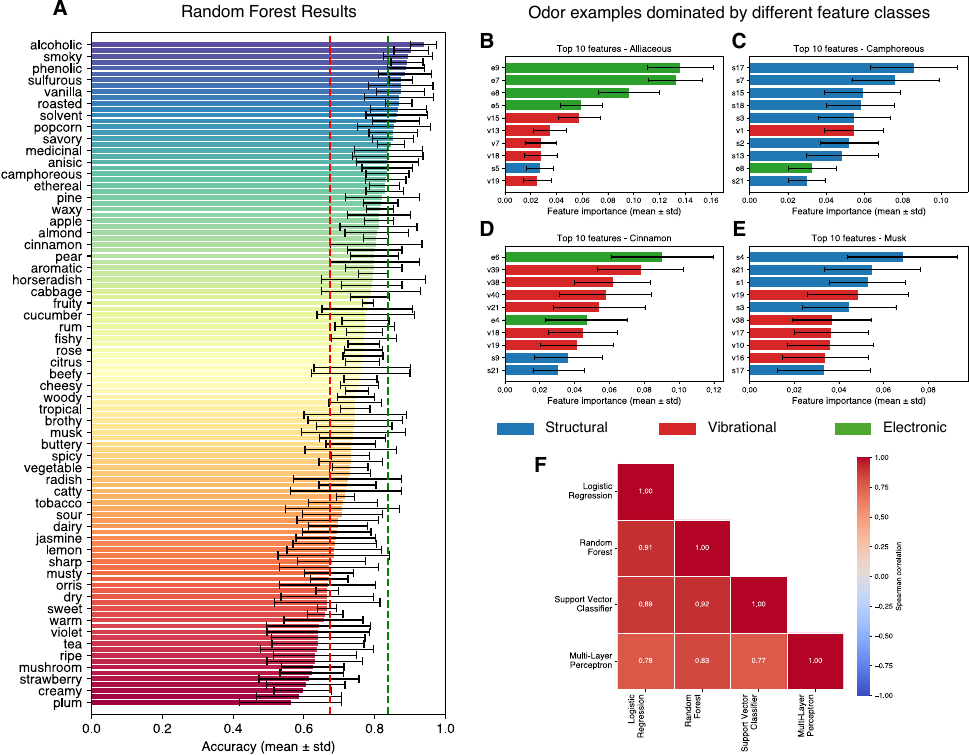}
    \caption{\label{fig:accuracy}
    ({\bf A}) Average accuracy of Random Forest models trained for individual odor categories. 
    The bars show the variance in the accuracy over the 100 random sampled balance of data. 
    Green and red dashed lines represent high- and low-accuracy thresholds, respectively. 
    ({\bf B}) panels highlighting the top ten molecular descriptors for the cinnamon, camphoreous, musk, and alliaceous odors, illustrating the combined relevance of vibrational, structural and electronic features.
    ({\bf C}) correlation of prediction between different machine learning algoritms: logistic regression (LR), random forest (RF), support vector classifier (SVC) and multilayer perceptron (MLP).
    }
\end{figure*}

Individual models trained for each odor exhibited accuracies ranging from moderate to near-perfect. Categories such as {catty}, {almond}, and {camphoreous} were predicted with high accuracy (above 0.8), whereas {tobacco}, {mushroom}, and {metallic} yielded lower scores (Fig.~\ref{fig:accuracy}A).
The inset panels highlight the ten most important molecular features for representative odor categories --- {cinnamon}, {camphoreous}, musk, and {alliaceous} --- illustrating how different physicochemical domains contribute to odor prediction, Fig.~\ref{fig:accuracy}B.
The {cinnamon} odor is predominantly influenced by vibrational descriptors ($v_j$), suggesting that characteristic vibrational modes may play a role in its recognition.
In contrast, {camphoreous} and {alliaceous} odors show stronger dependence on structural ($s_j$) and electronic ($e_j$) features, respectively, indicating that steric and charge-distribution effects govern their perceptual profiles.
Notably, the {musk} odor, exhibits a balanced contribution from vibrational ($v_{13}$, $v_{17}$, $v_{19}$), structural (rotational constants $s_1$-, $s_3$), and global shape descriptors [molecular weight ($s_4$), Kier Phi ($s_{21}$)], suggesting that muskiness emerges from the integrated influence of molecular flexibility and electronic distribution.
The results here predicted by the RF model is correlated with the prediction of other models, for instance, logistic regression (LR), support vector classifier (SVC), multilayer perceptron (MLP), indicating a robustness of our analysis, Fig.~\ref{fig:accuracy}C.

\begin{figure*}[th!]
    \centering
    \includegraphics[width=0.8\linewidth]{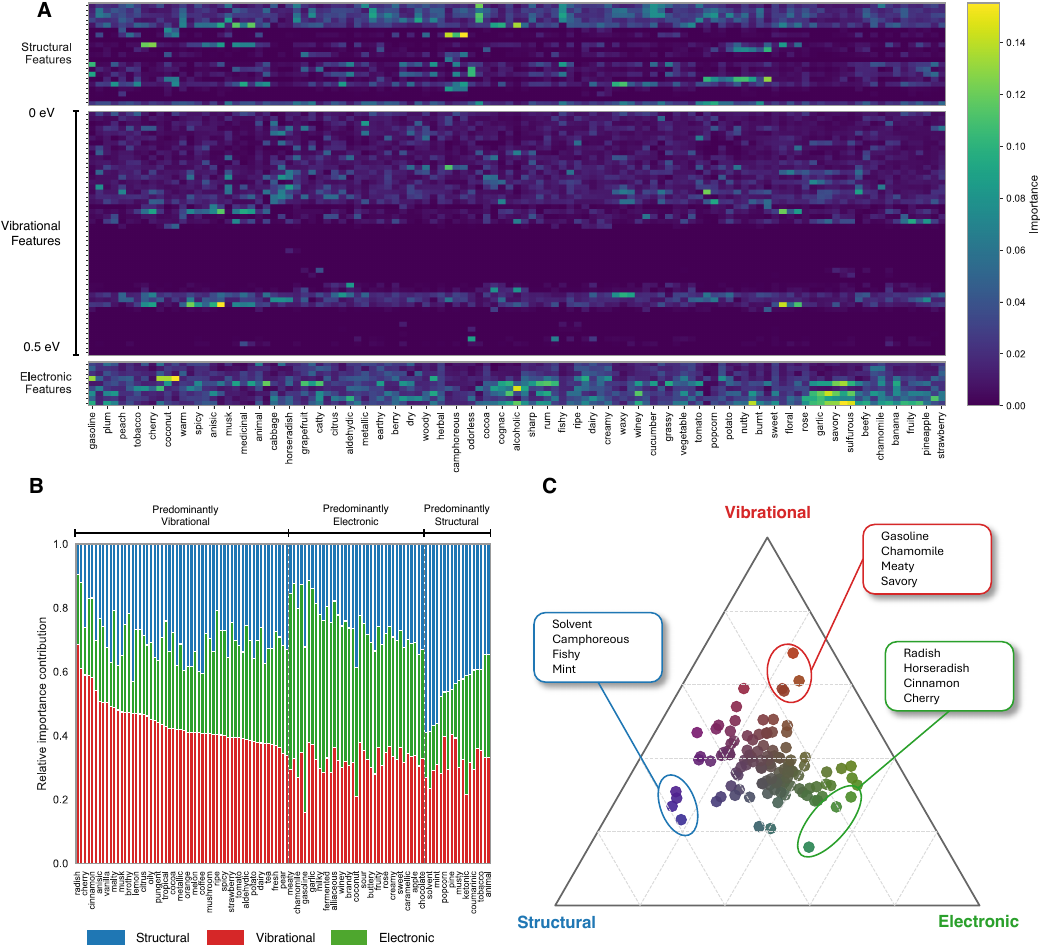}
    \caption{\label{fig:importance}
    ({\bf A}) Feature importance map obtained from Random Forest models trained for individual odor categories. 
    Each column represents an odor descriptor and each row corresponds to a molecular feature.
    Color intensity indicates the relative contribution of each feature to the prediction of a given odor, ranging from low (purple) to high (yellow) importance.
    All odors and molecular descriptors are included in the plot, for improved visualization only some labeled. 
    ({\bf B}) Relative contribution of electronic, vibrational, and structural descriptors to the prediction of each odor category. 
    Bars represent the summed feature importance within each molecular domain, ordered by cluster variance.
    ({\bf C}) Ternary diagram shows the odors with higher isolated mechanism importance.
}
\end{figure*}

A global view of feature importance across all odor classes is provided by the heatmap in Fig.~\ref{fig:importance}A.
Each odor is represented by a row and each descriptor by a column, with bright regions indicating strong feature importance.
The heatmap shows that no single descriptor dominates across all odor types.
Instead, distinct subsets of features drive specific perceptual domains: electronic descriptors associated with charge localization on heteroatoms dominate sulfurous and roasted odors, whereas structural descriptors such as the number of aliphatic rings and aromatic carbocycles are most relevant for mint and cherry, respectively.
This spatial segregation indicates that olfaction is a composite process shaped by coordinated contributions from multiple physicochemical domains.
Notably, the co-occurrence of vibrational, electronic, and structural features provides a unifying perspective that reconciles structure-based and vibration-based views of olfaction, and helps explain why topology-driven models can achieve high predictive performance \cite{SCIENCElee2023}.
Here, however, we emphasize interpretability and mechanistic insight rather than prediction accuracy.

To synthesize these trends at a higher level, molecular descriptors were grouped into three physicochemically defined domains—electronic, vibrational, and structural—and their summed importance was analyzed across odors (Fig.~\ref{fig:importance}B).
The summed feature importance within each domain quantifies the relative contribution of electronic, vibrational, and structural descriptors to odor prediction, offering a model-based view of the dominant physicochemical factors associated with each odor class.
Within this framework, electronic descriptors dominate fermented, winey, rum, gasoline, hay, garlic, meaty, and onion odors, whereas vibrational features are most prominent for radish, horseradish, cherry, cabbage, medicinal, balsamic, and musk categories.
Structural descriptors, in turn, predominate in camphoreous, solvent, mint, fishy, bread, popcorn, pine, and nutty odors.
Ordering odors by the variance of their domain-level contributions further separates cases in which multiple descriptor classes contribute comparably from those characterized by a clear dominance of a single physicochemical domain.
Because descriptors within each domain can be correlated, domain-level importance should be interpreted as a collective contribution rather than as evidence that individual features act independently.

These domain-level patterns are summarized in the ternary representation of Fig.~\ref{fig:importance}C, which organizes odor categories according to their relative electronic, vibrational, and structural contributions.
The resulting map highlights both distinct clusters and overlapping regions, indicating that odor perception spans partially shared physicochemical regimes rather than discrete mechanisms.
That is electronic, vibrational, and structural properties each modulate receptor interactions in distinct ways.
Electronic parameters could influence ligand–receptor affinity through charge distribution and polarizability; vibrational modes might contribute to frequency-specific coupling mechanisms; and structural descriptors likely govern steric complementarity and molecular flexibility.
While the present analysis does not establish mechanistic causality, it reveals that the olfactory space can be decomposed into interacting electronic, vibrational, and structural dimensions, each defining a distinct region of perceptual topology.

\subsection{Integration with Olfactory Receptor (OR) Data}

Although the analyses above reveal clear associations between electronic, vibrational, and structural molecular features and odor categories, these physicochemical contributions are not yet explicitly resolved at the level of olfactory receptors (ORs).
Particularly, receptor-level discussions of vibrational effects have been conceptual rather than molecular specific \cite{PRLbrookes2007}, despite empirical evidence \cite{PNASfranco2011}. 
While the dynamical aspect of structural molecular feature are known for specific OR \cite{billesbolle2023structural}.
Here, rather than attempting to establish a microscopic tunneling mechanism, we take a complementary and data-driven approach.
We integrate the physicochemical trends identified above with a dataset describing the human olfactory receptor, with the aim of assessing whether specific electronic, vibrational, or structural signatures preferentially associate with subsets of ORs.
This integration provides an empirical link between molecular descriptors and receptor-level organization, and allows us to evaluate whether the physicochemical domains highlighted in Fig.~\ref{fig:importance} are reflected in the structure and diversity of the human olfactory system.

\begin{figure*}[th!]
    \centering
    \includegraphics[width=0.8\linewidth]{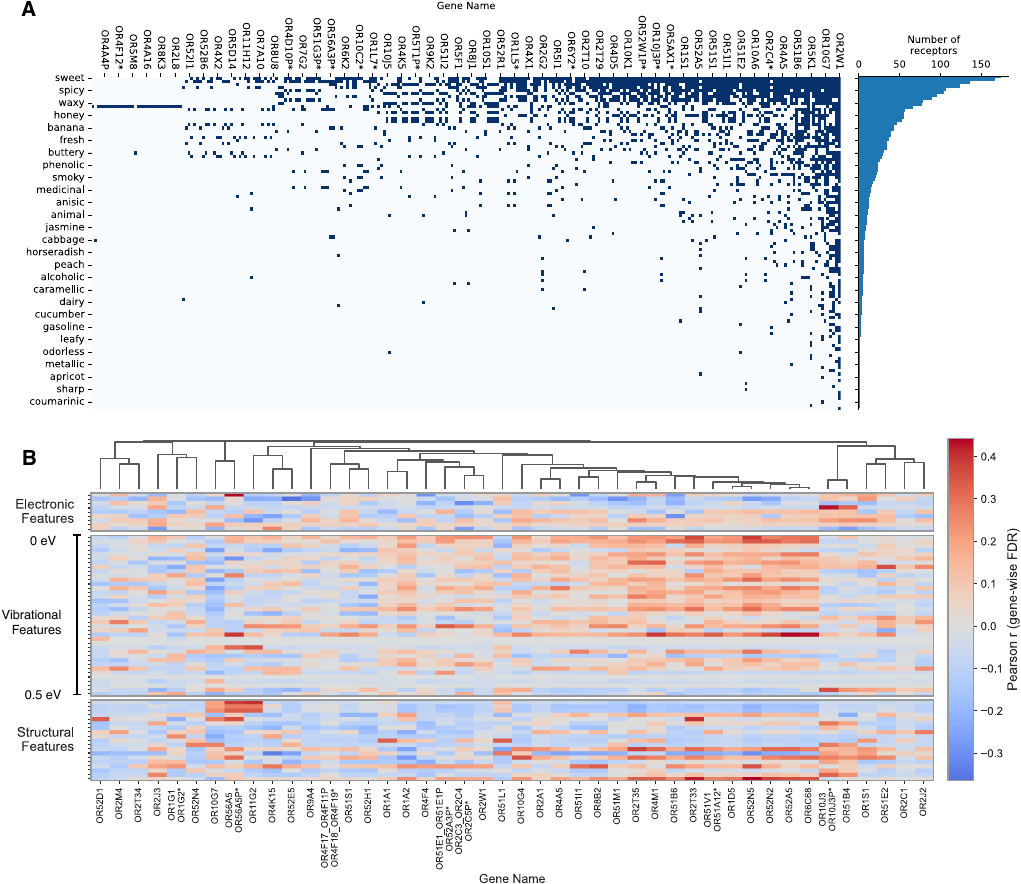}
    \caption{\label{fig:bio1}
    ({\bf A}) Display of binary receptor–odor associations (responsive = 1) for monomolecular odorants. 
    Each row corresponds to a human olfactory receptor (OR gene), and each column to an odor descriptor.
    ({\bf B}) Correlation between OR activation and associated molecular features. 
    }
\end{figure*}

The integration of the M2OR database with our predicted odor-molecule associations supports the classical combinatorial model of olfactory coding \cite{malnic1999combinatorial, saito2009odor}.
As shown in Fig.~\ref{fig:bio1}A, individual olfactory receptors exhibit distinct yet partially overlapping tuning profiles across odor categories.
In this framework, OR2J3, OR2M4, and OR5A1 show preferential responsiveness to floral, fruity, and fatty odorants, respectively, whereas OR51E1 and OR52E4 display broader activation patterns that extend toward sulfurous and woody compounds, highlighting the diversity and redundancy inherent to combinatorial odor coding.

Consistent with these predictions, OR2J3 has been experimentally shown to respond to cis-3-hexen-1-ol, a compound associated with a characteristic ``green'' or ``cut grass'' odor \cite{mcrae2012genetic}.
However, we also observe discrepancies between our predictions and existing experimental evidence.
For example, although OR5A1 emerged in our analysis as being associated with fatty-like odorants, experimental studies to date have primarily linked this receptor to floral or fragrant compounds \cite{jaeger2013mendelian}.
Such inconsistencies should be interpreted in the context of the limited functional annotation of the human olfactory receptor repertoire.
Despite the remarkable discriminatory capacity of the human olfactory system, ligands have been experimentally identified for only $10\%$ of the approximately $400$ intact human olfactory receptors, leaving the vast majority functionally orphaned \cite{mainland2015human}.
This scarcity of receptor–ligand mappings represents a major bottleneck in the field and constrains direct inference between receptor activation patterns and perceptual odor qualities.

Overall, our results suggest that the proposed predictive framework captures core aspects of combinatorial olfactory coding, while also exposing clear gaps between computational predictions and the current experimental record.
The concordance observed for receptors such as OR2J3 lends confidence to the approach, whereas cases like OR5A1 highlight how limited functional data still shape, and sometimes constrain, interpretation.
In this sense, mismatches between predictions and published ligand assignments likely reflect not only limitations in the model, but also the sparse and uneven coverage of receptor–ligand space.
Given that most human olfactory receptors remain functionally uncharacterized, integrative analyses such as ours can help identify promising receptor–odorant relationships and guide future experimental efforts toward a more complete view of olfactory coding.

The hierarchical clustering of olfactory receptors based on molecular descriptor similarity (Fig.~\ref{fig:bio1}B) revealed clear and organized associations between receptor response profiles and major physicochemical domains—electronic, vibrational, and structural—indicating that olfactory receptors group according to shared physicochemical sensitivities.
Some receptors, such as OR5A1, OR1A1, and OR2J3, showed stronger correlations (red) with electronic and low-frequency vibrational parameters, whereas others, including OR52E1 and OR51E1, were more related to structural descriptors linked to molecular topology and flexibility.
This pattern suggests that subsets of receptors may preferentially respond to odorants defined by distinct physicochemical properties, pointing toward a multidimensional coding scheme that combines electronic, vibrational, and structural information. 
At the highest-confidence threshold (FDR $q<5\%$, Pearson $r>0.4$), a limited set of receptor-descriptor associations emerges. OR56A5 and OR10J3 preferentially correlate with electronic ($e_1$, $e_4$) and vibrational ($v_{24}$, $v_{47}$) features; OR52A5 and OR6C68 show dominant vibrational correlations ($v_{24}$); and OR52N5 and OR2T33 are more strongly associated with structural ($s_{21}$) and vibrational ($v_2$, $v_1$) descriptors.
These patterns reinforce the view that subsets of receptors may exhibit preferential sensitivity to distinct physicochemical regimes rather than sharing a uniform selectivity profile.

Although these clusters resemble those obtained in our feature-importance analyses, it remains uncertain to what extent receptor selectivity truly aligns with the predicted physicochemical domains --- electronic, vibrational, or structural.
These preliminary associations suggest a structured, potentially multi-level organization of odor coding; however, further experimental validation will be necessary to determine whether the computationally derived molecular features correspond directly to receptor-level activation within the human olfactory system.

\section{Conclusion} 

Here, we investigated how molecular structural, electronic, and vibrational properties relate to odor perception by integrating first-principles molecular descriptors, perceptual odor annotations, and olfactory receptor interaction data.
Across a diverse set of nearly 3,500 odorant molecules, our analyses show that odor categories are not associated with single physicochemical determinants, but instead reflect patterns distributed across multiple molecular domains.

Our results show that correlation analyses link specific odor labels to well-defined feature ranges, including narrow vibrational energy intervals and structural descriptors related to hydrogen bonding, heteroatom content, hydrophobicity, and molecular flexibility.
Random Forest models extend these findings by capturing non-linear relationships and revealing marked differences in predictability across odors.
Together, these analyses show that the molecular features relevant for odor classification vary across odor categories, with no single signature shared across all odors.

We also found that feature-importance analyses reveal how these molecular patterns are organized, with different odor categories relying on distinct combinations of electronic, vibrational, and shape-related features.
As a result, odor space is non-uniform, with different odors occupying distinct regions of molecular feature space.

Integration with olfactory receptor data is consistent with a combinatorial coding scheme, in which individual receptors respond to multiple odor categories.
Where experimental data are available, several predicted receptor–odor associations agree with known ligands, while remaining mismatches mainly reflect the limited functional annotation of the human olfactory receptor repertoire.
We also reinforce that biases intrinsic to machine-learning predictions, arising from data imbalance and limited receptor–ligand coverage, cannot be fully excluded.

In summary, our findings place quantitative constraints on models of olfaction, indicating that odor perception is better described by heterogeneous, odor-specific molecular determinants than by a universal encoding scheme.
This perspective provides a practical basis for organizing odor space and for prioritizing molecular features and receptor–odorant pairs for future experimental investigation.

\section*{Acknowledgments}

The authors acknowledge financial support from the Brazilian agencies FAPESP (grants 23/09820-2, 24/00989-7 and 2024/22392-2), CNPq (INCT - Materials Informatics, projects 422069/2023-0 and 313301/2025-5), LNCC (Laboratório Nacional de Computação Científica) (Project didmat).

\section*{Author contributions:} 
F. Crasto de Lima performed the DFT calculations and, together with L. N. Lemos, wrote the manuscript. L. N. Lemos carried out the olfactory receptor analysis. P. Zanineli and G. R. Schleder developed the machine-learning models. A. Fazzio conceived, supervised, and funded the project. All authors contributed to the interpretation of the results and discussions.

\section*{Competing interests}
Authors declare that they have no competing interests.

\section*{Data and materials availability}
All data are available in the main text or the supplementary materials.

\bibliography{refs}% Produces the bibliography via BibTeX.

\clearpage
%\appendix
\setcounter{section}{0}
\setcounter{figure}{0}
\setcounter{table}{0}
\renewcommand{\thesection}{S\arabic{section}}
\renewcommand{\thefigure}{S\arabic{figure}}
\renewcommand{\thetable}{S\arabic{table}}

{\Large \centering \bf SUPPLEMENTAL MATERIALS}
\begin{center}
{\bf \large Heterogeneous molecular determinants of odor perception} \\
P. Zanineli; E. V. C. Lopes; G. R. Schleder; L. N. Lemos; F. Crasto de Lima and A. Fazzio
\end{center}

\section{Leffingwell database}

\begin{figure}[h!]
    \centering
    \includegraphics[width=1\linewidth]{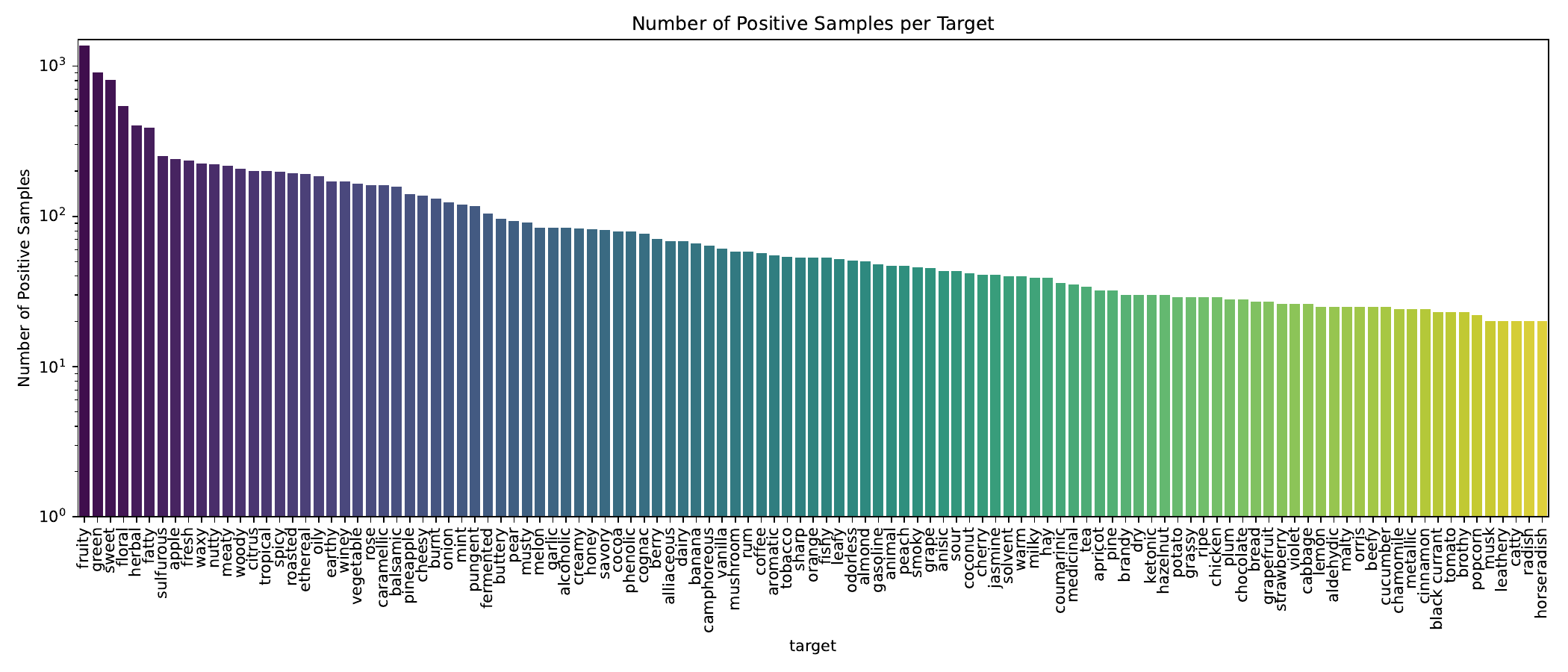}
    \caption{\label{fig:s-num-data-odor} Number of data points (molecules) for each of the odors.}
\end{figure}

\begin{figure}[h!]
    \centering
    \includegraphics[width=1\linewidth]{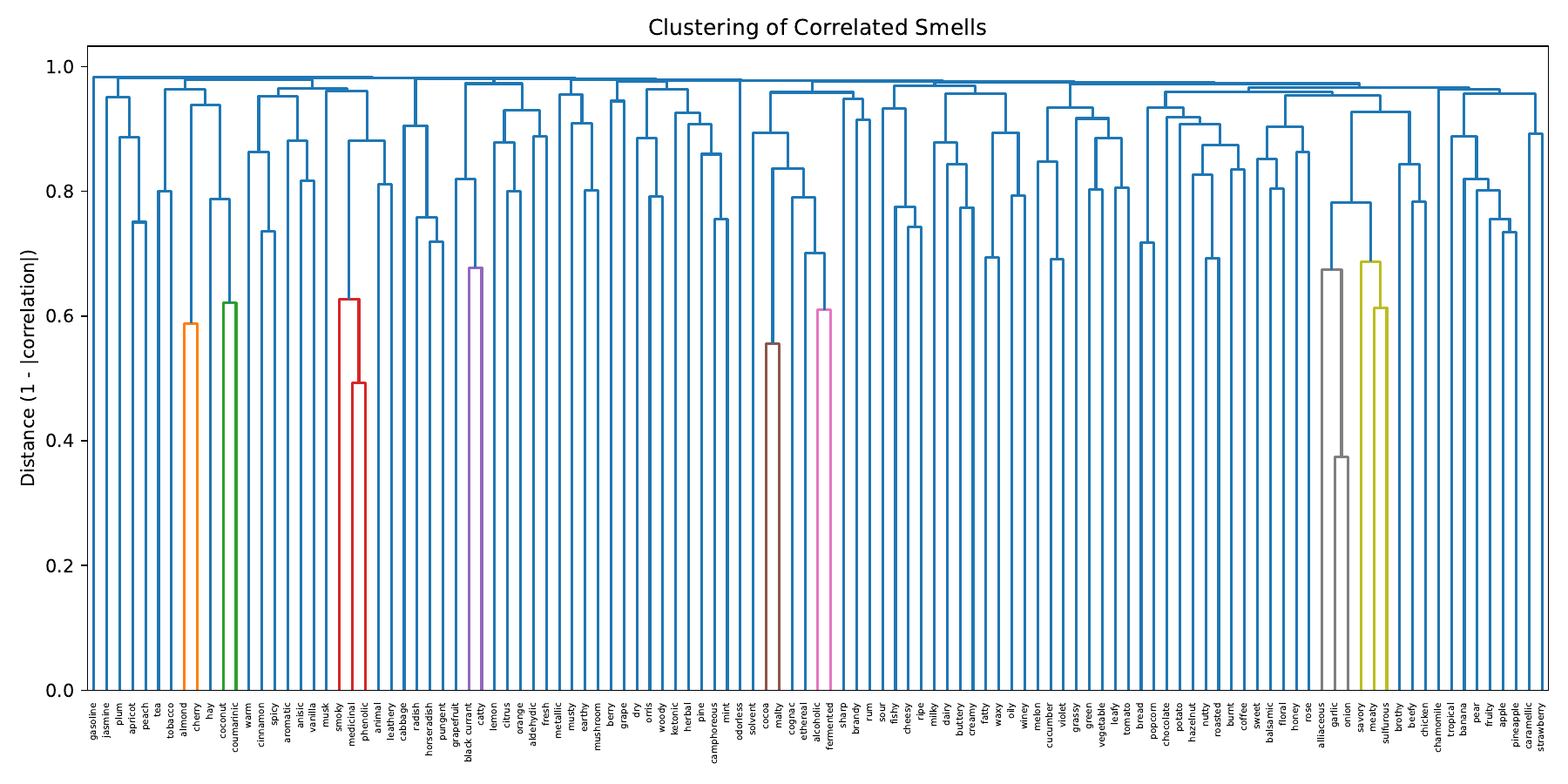}
    \caption{\label{fig:s-dendogram-odor} Correlation distance between different odors.}
\end{figure}

In the Table~\ref{tab:odors} we show the name of each odor present in the Leffingwell database. In Fig.~\ref{fig:s-num-data-odor} we show the number of data/molecules manifesting each of the odors. Here se see an imbalance of the dataset with some odors containing thousands of data while others within tens of data points. Additionally, we have computed the proximity of those odors by the correlation distance
\begin{equation}
    d = 1 - r_{xy},
\end{equation}
where $r_{xy}$ is the Pearson correlation measuring the average co-occurrence of odors among the molecules. As shown in Fig.~\ref{fig:s-dendogram-odor}, there is a significant distance between the odors guarantying that they are nonequivalent. However, the hierarchical representation allow to infer a closeness between the groups with distance $d<0.7$.

\begin{table}[h!]
\begin{ruledtabular}
\caption{\label{tab:odors} List of odors present in the Leffingwell database.}
\begin{tabular}{lllll}

\multicolumn{5}{c}{Odors names} \\
\hline
alcoholic&aldehydic&alliaceous&almond&animal\\
anisic&apple&apricot&aromatic&balsamic\\
banana&beefy&berry&black&currant\\
brandy&bread&brothy&burnt&buttery\\
cabbage&camphoreous&caramellic&catty&chamomile\\
cheesy&cherry&chicken&chocolate&cinnamon\\
citrus&cocoa&coconut&coffee&cognac\\
coumarinic&creamy&cucumber&dairy&dry\\
earthy&ethereal&fatty&fermented&fishy\\
floral&fresh&fruity&garlic&gasoline\\
grape&grapefruit&grassy&green&hay\\
hazelnut&herbal&honey&horseradish&jasmine\\
ketonic&leafy&leathery&lemon&malty\\
meaty&medicinal&melon&metallic&milky\\
mint&mushroom&musk&musty&nutty\\
odorless&oily&onion&orange&orris\\
peach&pear&phenolic&pine&pineapple\\
plum&popcorn&potato&pungent&radish\\
ripe&roasted&rose&rum&savory\\
sharp&smoky&solvent&sour&spicy\\
strawberry&sulfurous&sweet&tea&tobacco\\
tomato&tropical&vanilla&vegetable&violet\\
warm&waxy&winey&woody & 
\end{tabular}
\end{ruledtabular}
\end{table}

\section{Average Correlations}

\begin{figure}[h!]
    \centering
    \includegraphics[width=1\linewidth]{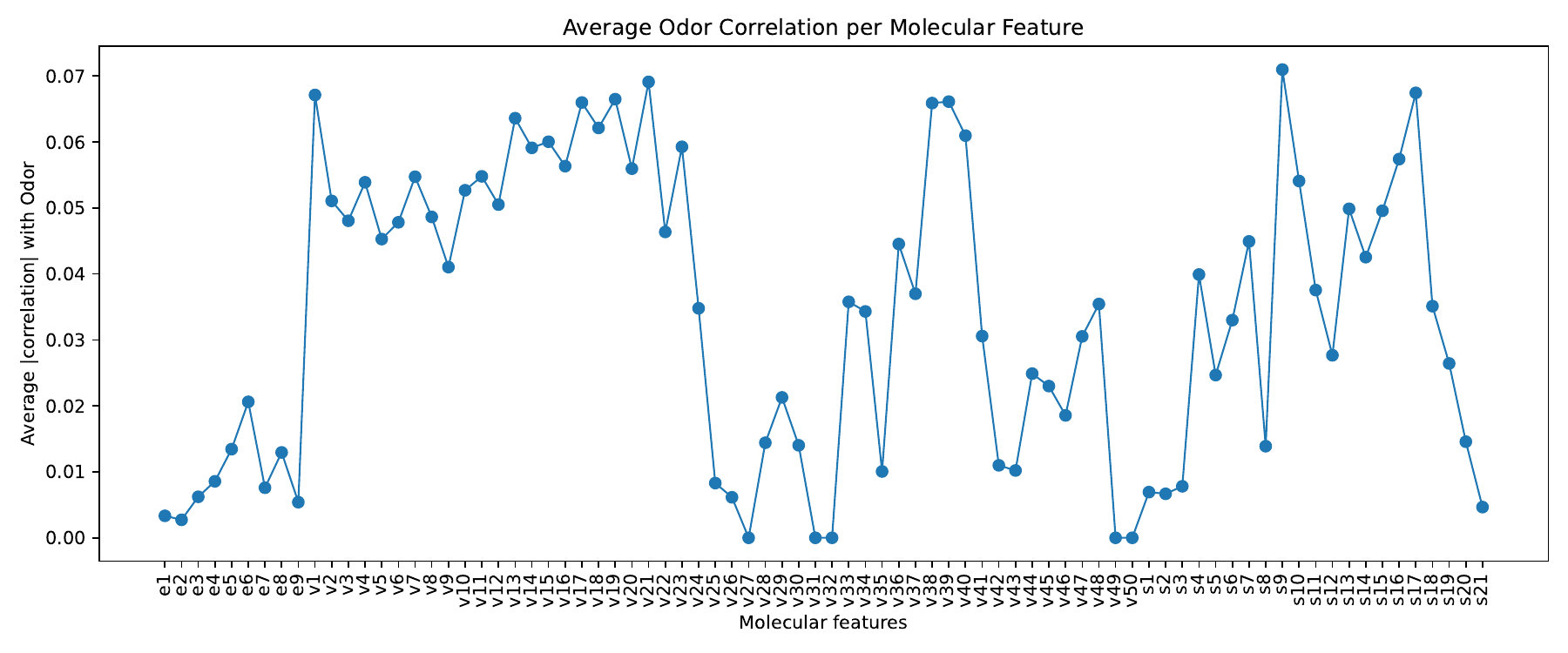}
    \caption{\label{fig:s-avg-corre} Absolute correlation of the molecular features averaged over all odors.}
\end{figure}

Taking Pearson correlation shown in Fig~\ref{fig:correlation}, and averaging it absolute over all odors we can see the relation shown in Fig.~\ref{fig:s-avg-corre}. Here we can see that in average the electronic features have lower correlation than the vibrational and structural features. Particularly, the higher correlation in each domain is (i) for the electronic features the e$6$ i.e. the LUMO value; (ii) for the vibrational features the v1 up to v23 which is related to vibrational modes with energy between $0$ and $0.23$\,eV, but also the interval between v38 to v40, which is the energy interval between $0.38$ to $0.40$\,eV; (iii) for the structural features the s9, s10, s16 and s17 being respectively the Aromatic carbocycles,  Aromatic heterocycles,  Heterocycles and Rotatable bonds count.

\section{Robustness of different algorithms trained for individual odors}

\begin{figure}[h!]
    \centering
    \includegraphics[width=.75\linewidth]{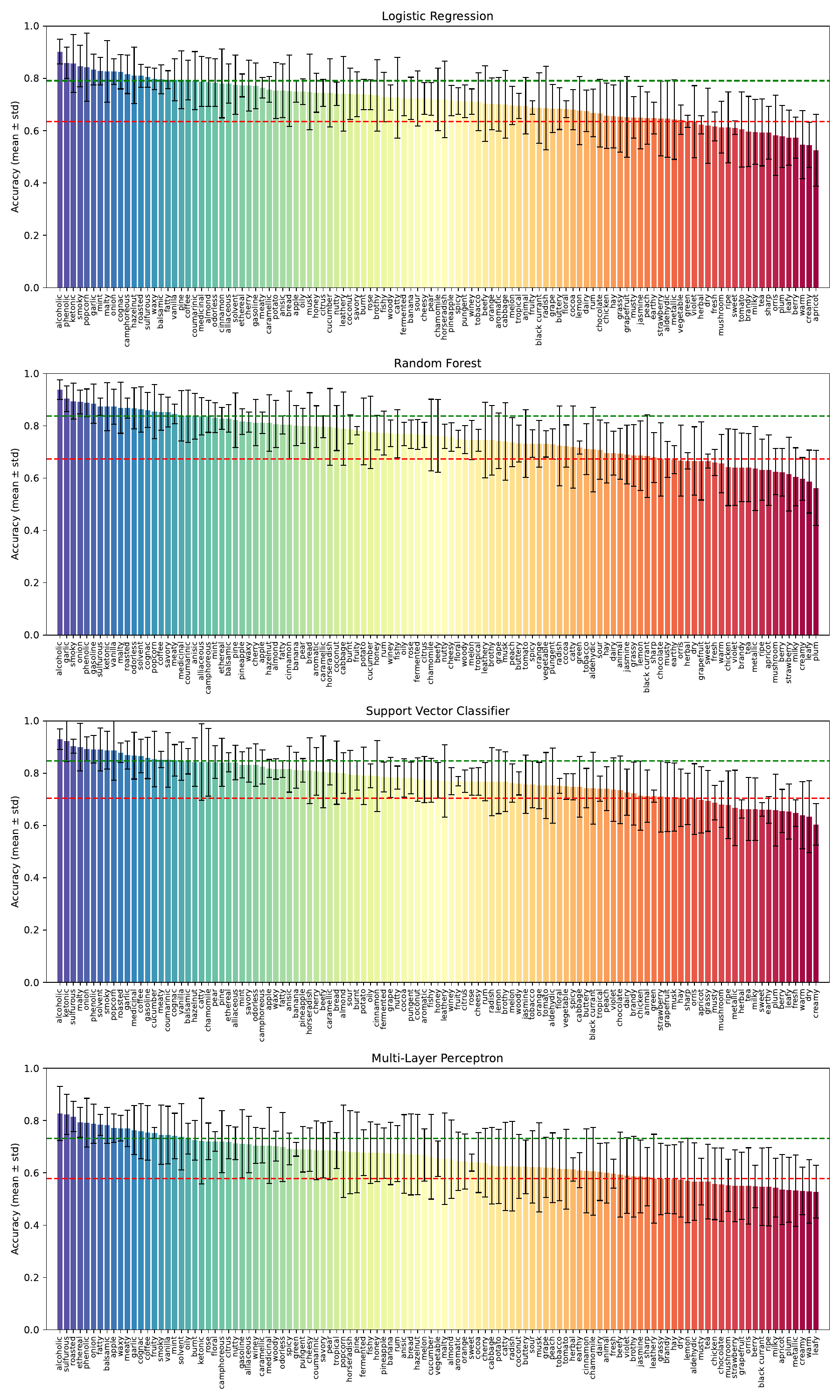}
    \caption{Per-odor accuracies obtained by different machine learning classification methods. From the first to the last panel, it is shown the Logistic Regression, Random Forest, Support Vector Machine, and Multi-Layer Perceptron models.}
    \label{fig:different_algos}
\end{figure}

To evaluate how consistently different machine learning models can recognize specific odor categories, we compared the per-odor accuracies obtained by four representative algorithms: Logistic Regression, Random Forest, Support Vector Classifier (SVC), and a Multi-Layer Perceptron (MLP). These results are summarized in the composite plot in Figure \ref{fig:different_algos}.

Across all models, odors with stronger, chemically distinct signatures — such as alcoholic, phenolic, roasted, sulfurous, and garlic — tend to achieve consistently high accuracies, often exceeding 0.8. These categories exhibit low variance across cross-validation folds, indicating that the underlying molecular descriptors reliably separate them from the rest of the odor space.

Conversely, more diffuse or perceptually complex odor descriptors—such as creamy, apricot, mushroom, berry, and plum—show substantially lower performance, with accuracies frequently below 0.5 and higher standard deviations. This suggests overlapping chemical features that challenge even non-linear models.

Among the models, Random Forest and SVC demonstrate the most stable overall behavior, maintaining higher median accuracies across the full odor spectrum. Logistic Regression, despite its linear nature, performs competitively on odors whose molecular features are well aligned with linear separability, but struggles considerably with subtler classes. The MLP shows slightly improved performance over Logistic Regression for ambiguous odors but exhibits larger variance, implying sensitivity to optimization and sample imbalance.

\begin{figure}[h!]
    \centering
    \includegraphics[width=.9\linewidth]{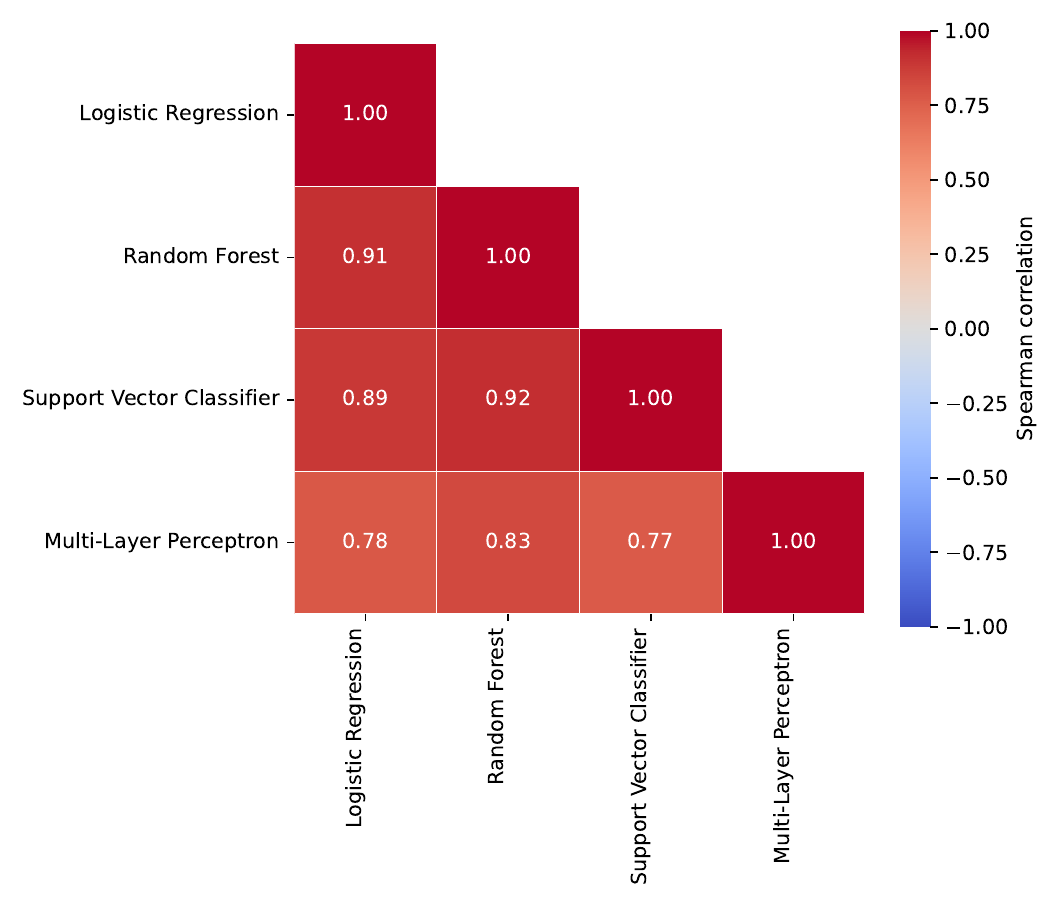}
    \caption{Heatmap of the Pearson correlation between per-odor prediction accuracies across the four evaluated algorithms. Darker red cells indicate stronger positive correlations, meaning the models tend to agree on which odors are easier or harder to classify.}
    \label{fig:model_pearson}
\end{figure}

Taken together, these findings reveal that odor prediction is governed by a highly structured yet uneven landscape of chemical–perceptual relationships. While individual odor classes differ markedly in difficulty, the four learning algorithms tend to agree on which odors are intrinsically easier or harder to classify. This is quantitatively supported by the strong Spearman correlations observed between their per-odor accuracy profiles in Figure \ref{fig:model_pearson}.

Logistic Regression, Random Forest, and SVC exhibit particularly high mutual correlations ($\rho \approx 0.89–0.92$), indicating a shared underlying ranking of odor discriminability despite their architectural differences. Even the MLP—which shows greater variance and occasional drops in performance—remains moderately aligned with the other models ($\rho \approx 0.75–0.83$). The convergence of these models toward similar error patterns suggests that the primary bottleneck is not algorithmic choice, but rather the intrinsic ambiguity of certain odor categories and the limits imposed by the currently available molecular descriptors.

\begin{figure}[h!]
    \centering
    \includegraphics[width=1\linewidth]{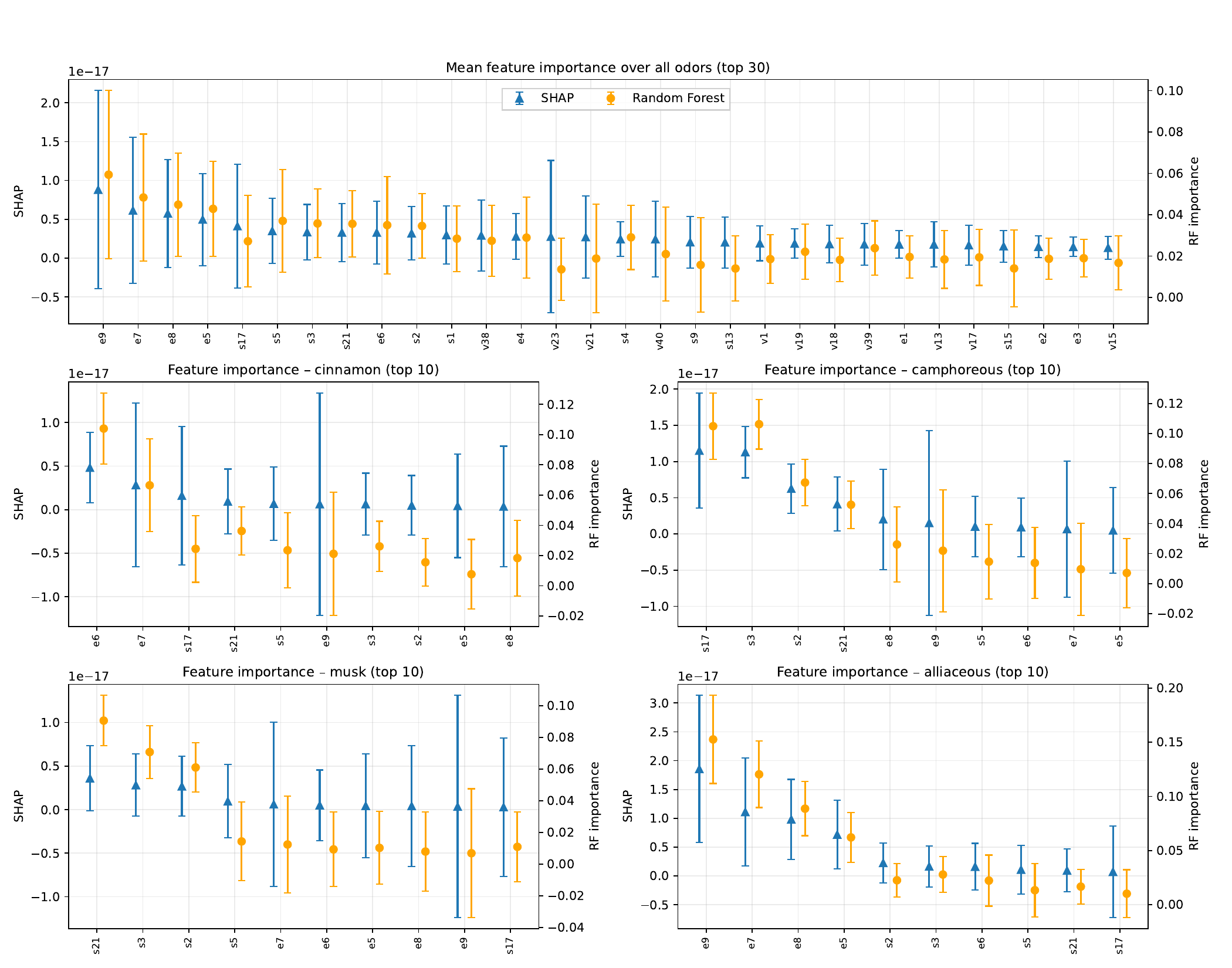}
    \caption{Top panel: Global mean SHAP values and RF feature importances computed across all 113 odors for the top 30 features, with error bars representing the standard deviation over all odors. Bottom panels: Odor-specific comparison of SHAP and RF importance for the top 10 features of four representative odors (cinnamon, camphoreous, musk, and alliaceous), plotted with global standard-deviation error bars to contextualize variability.}
    \label{fig:feature_comparison}
\end{figure}

Building on this overall convergence across models, we next examine whether their internal attributions also display similar patterns. Although SHAP values are derived from the trained Random Forest model and therefore not fully independent from RF impurity-based importance, the observed agreement across aggregation levels suggests a robust ranking of influential descriptors rather than an artifact of a single attribution metric, Figure \ref{fig:feature_comparison}. While RF feature importance provides a coarse estimate of each descriptor's contribution, SHAP offers a more principled, locally faithful, and model-agnostic quantification of feature effects, enabling a deeper understanding of how specific molecular properties influence odor predictions.

In the top panel, we show the global mean importance values across all 113 odors for the 30 most relevant features. Both SHAP and RF exhibit similar overall trends, indicating that the two methods capture convergent structural determinants influencing model performance. However, the error bars—representing the standard deviation computed over all odors—highlight substantial variability for several features, suggesting that certain descriptors have highly odor-dependent contributions. It is worth pointing out that the absolute scale of SHAP values depends on the numerical range of the model output and feature normalization; therefore, our interpretation focuses on relative magnitudes and feature rankings rather than absolute values.

This odor specificity becomes even more apparent in the four bottom panels, which detail the top 10 features for the odors cinnamon, camphoreous, musk, and alliaceous. Although these odors share some of the most influential descriptors, their relative ranking and magnitude differ markedly between SHAP and RF, reflecting the inherently selective nature of odorant–receptor interactions represented in the dataset.

\begin{figure}
    \centering
    \includegraphics[width=1\linewidth]{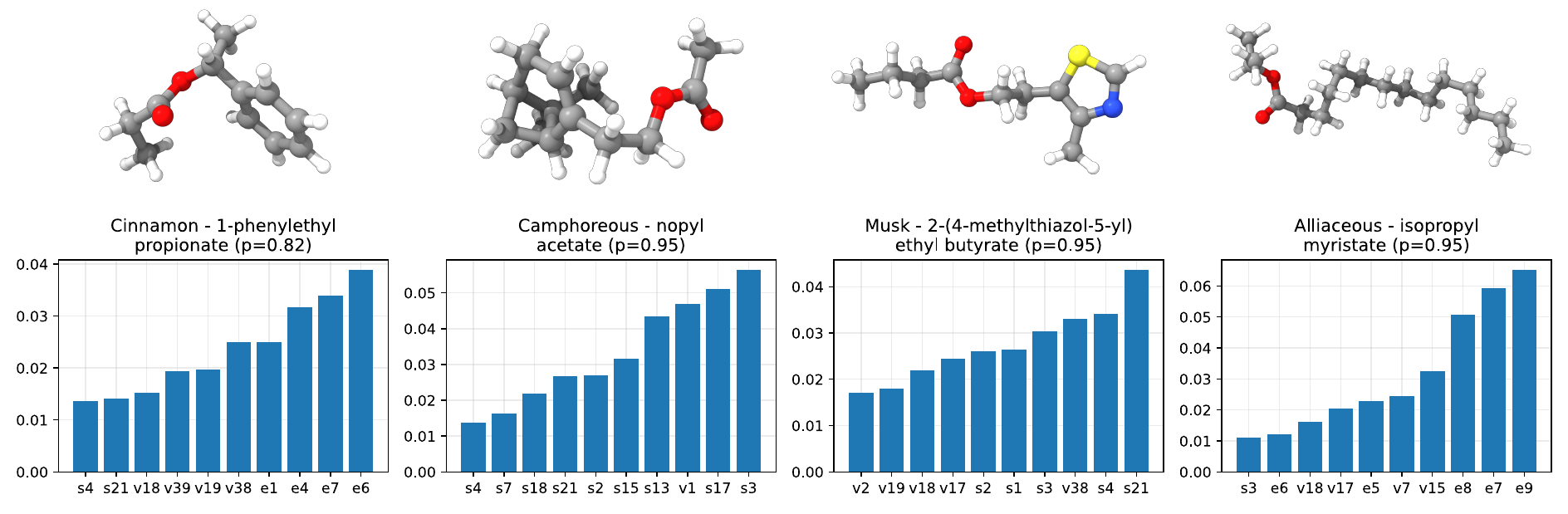}
    \caption{Top 10 features for a given molecule to properly predict the presence of a given odor. From left to right, cinnamon, camphoreous, musk, and alliaceous are shown.}
    \label{fig:rep_mols}
\end{figure}

Extending our analysis to the properties most relevant for a given molecule, we examine a set of representative odors, as illustrated in Figure \ref{fig:rep_mols} from the SHAP calculation. This allows us to further assess the consistency between the predicted important features and the known odor-characteristic properties, thereby establishing clearer connections between molecular structure and olfactory behavior.

We selected four molecules associated with distinct odor families—cinnamon, camphoreous, musk, and alliaceous—chosen for their contrasting chemical profiles. Notably, 1-phenylethyl propionate and isopropyl myristate exhibit predominantly electronic features among their most influential predictors, whereas nopyl acetate and 2-(4-methylthiazol-5-yl)ethyl butyrate are characterized by structural descriptors as their principal contributions. This contrast highlights how different odor classes rely on distinct molecular determinants within the learned feature representation, reinforcing the interplay between electronic and structural factors in shaping olfactory perception.

%%%% Comentar o bloco abaixo para remover o plot de todos os odores
%\newcommand{\figpath}{s}
%\newcommand{\figext}{pdf}
%\foreach \i in {8,9,10,11,12,13,14,15,16,17,18,19,20,21,22}{
%    \IfFileExists{\figpath/\i.\figext}{
%        \begin{figure}[h!]
%            \centering
%            \includegraphics[width=\linewidth]{\figpath/\i.\figext}
%            \caption{Odors comparison}% - \i}
%        \end{figure}
%    }{}
%}
%%%%

\newpage

\begin{figure}
            \centering
            \includegraphics[width=\linewidth]{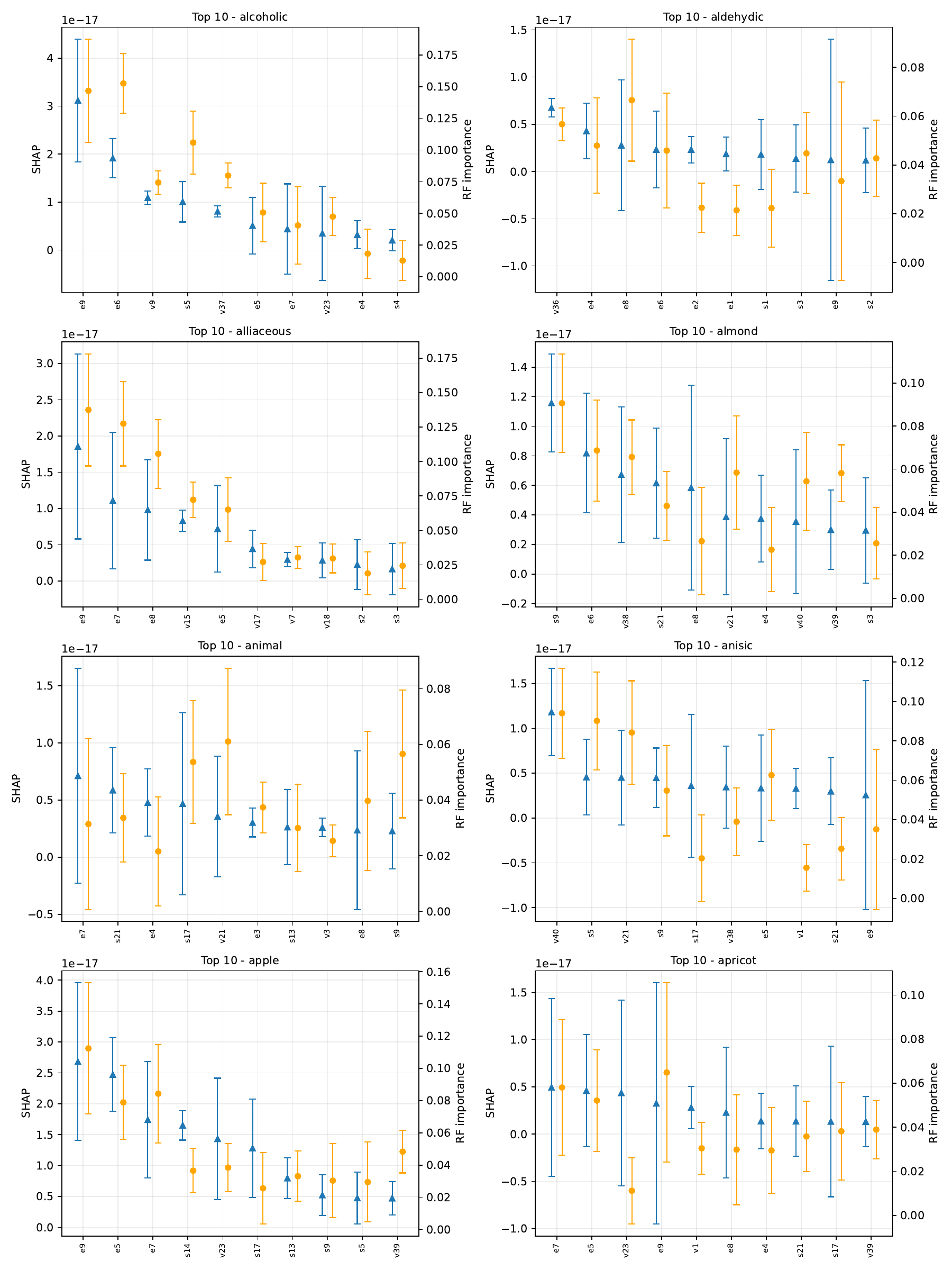}
            \caption{Odors comparison}
\end{figure}

\newpage

\begin{figure}
            \centering
            \includegraphics[width=\linewidth]{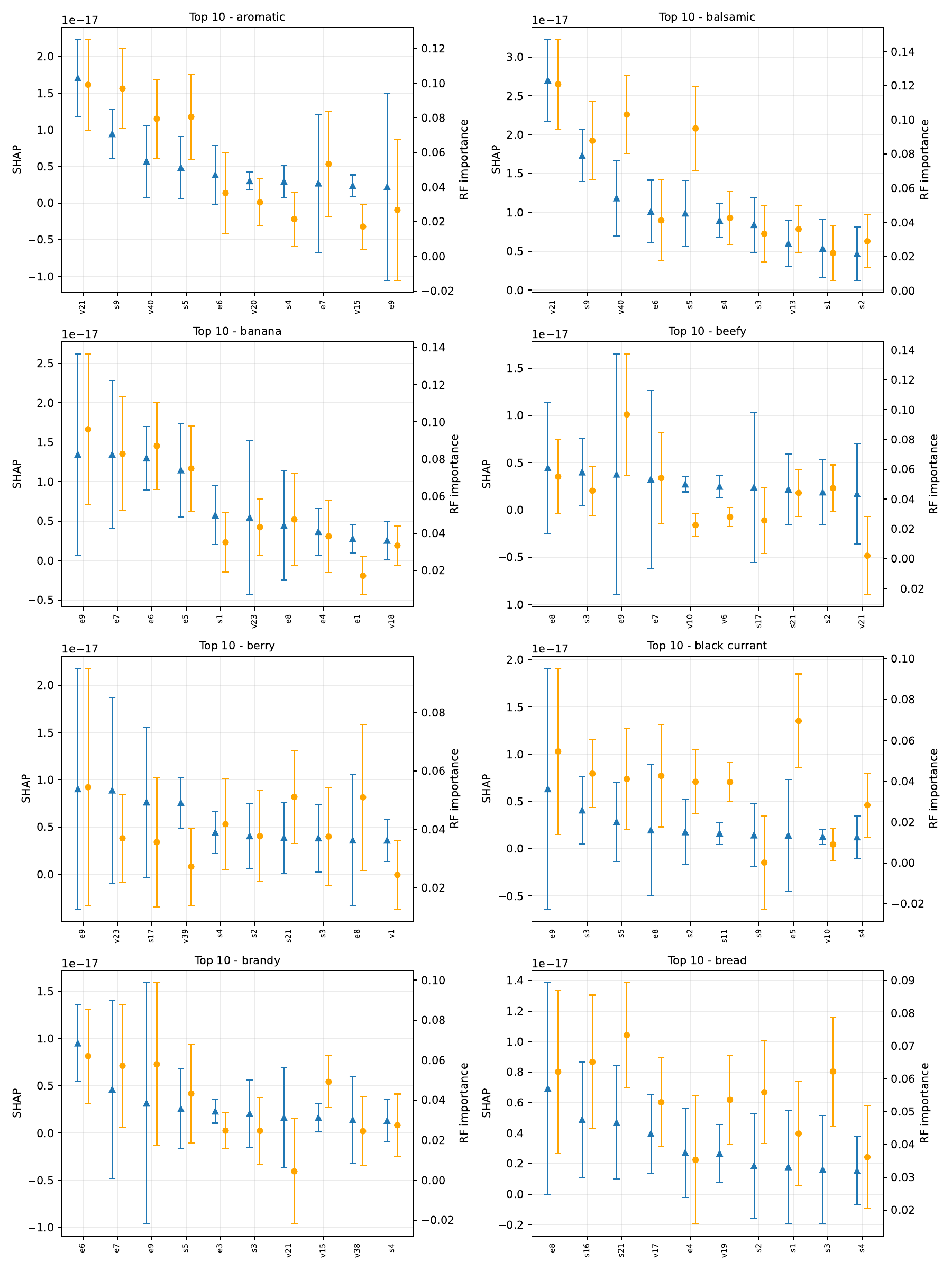}
            \caption{Odors comparison}
\end{figure}

\newpage

\begin{figure}
            \centering
            \includegraphics[width=\linewidth]{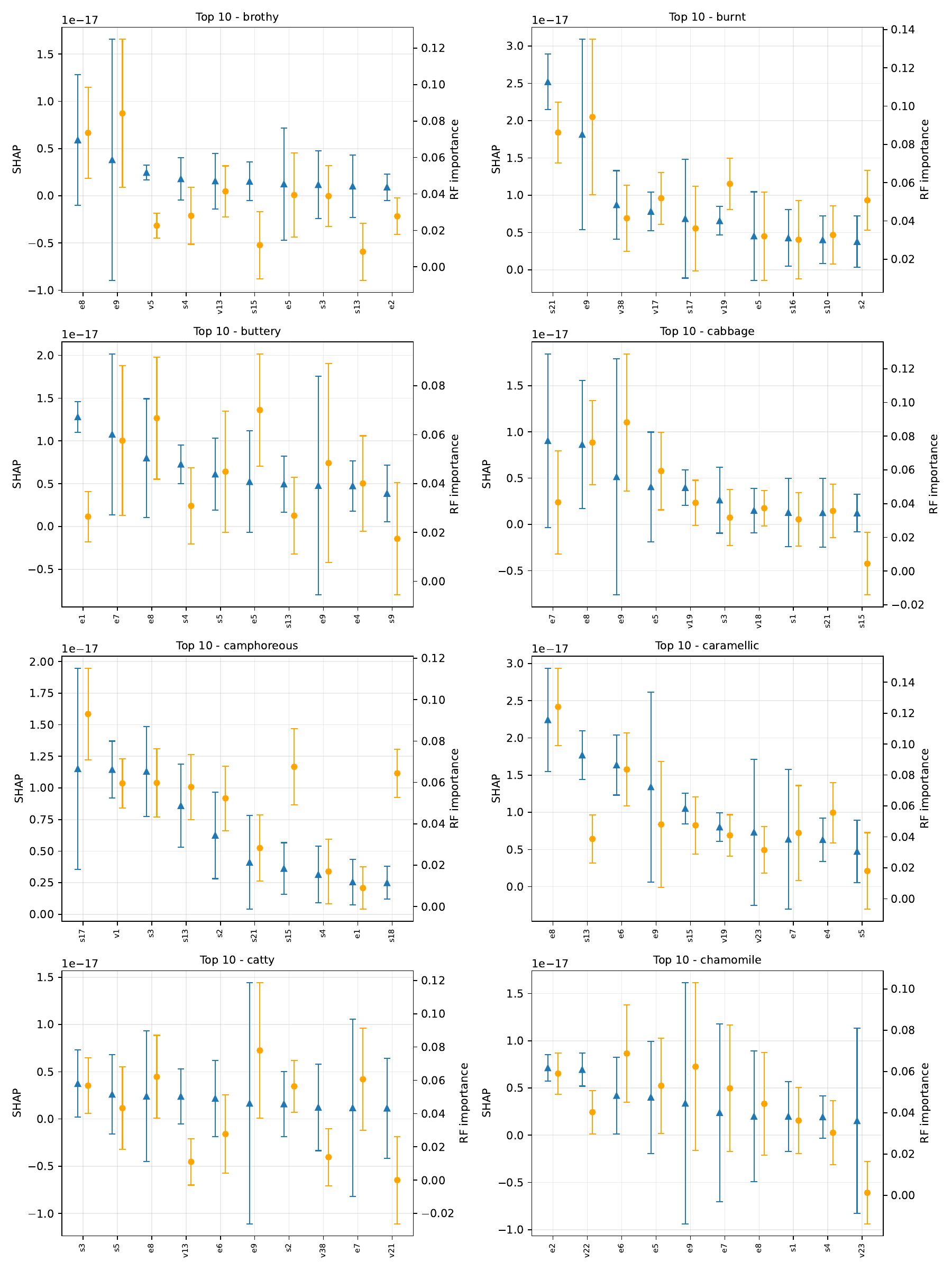}
            \caption{Odors comparison}
\end{figure}

\newpage

\begin{figure}
            \centering
            \includegraphics[width=\linewidth]{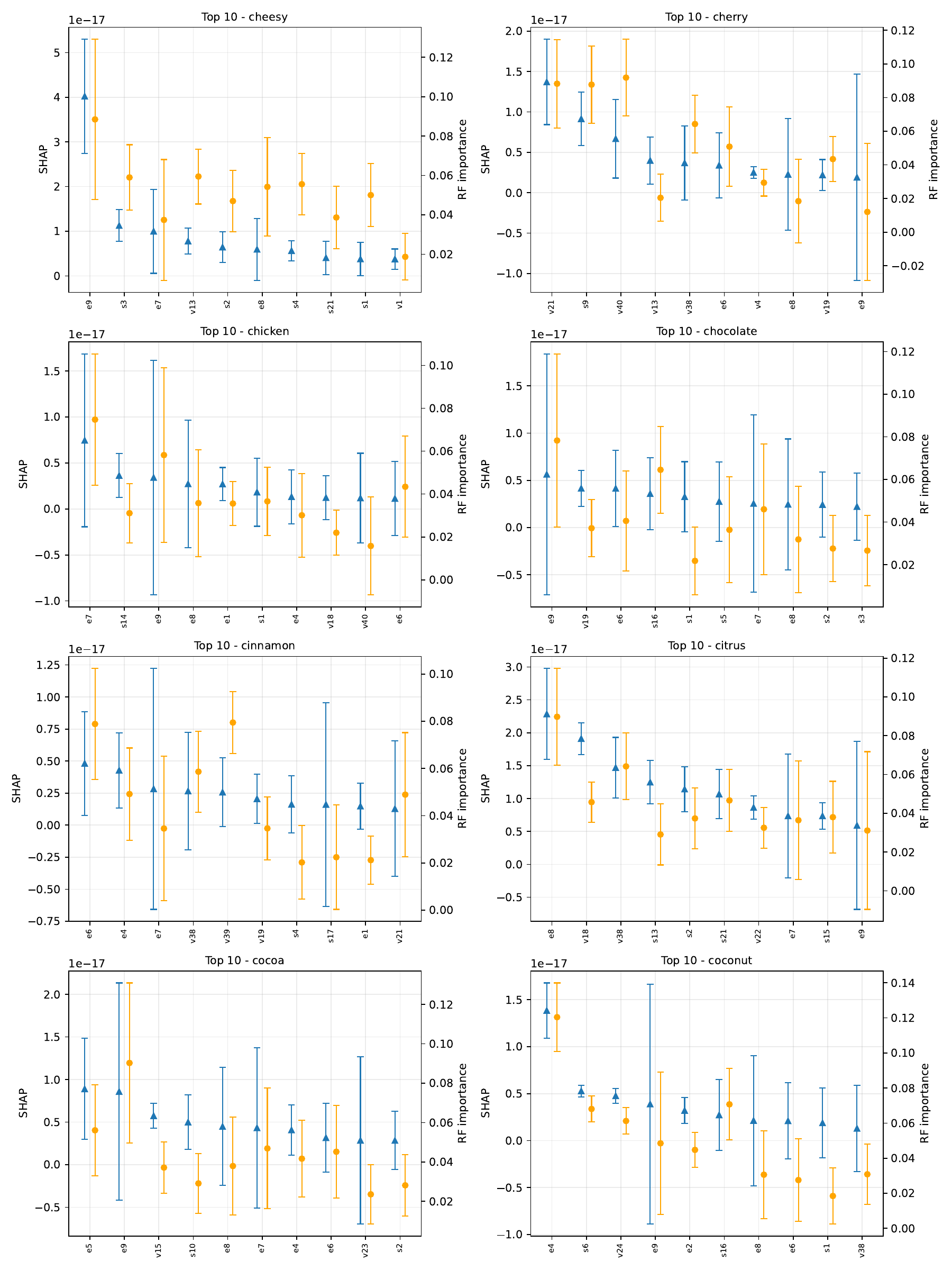}
            \caption{Odors comparison}
\end{figure}
        
\newpage

\begin{figure}
            \centering
            \includegraphics[width=\linewidth]{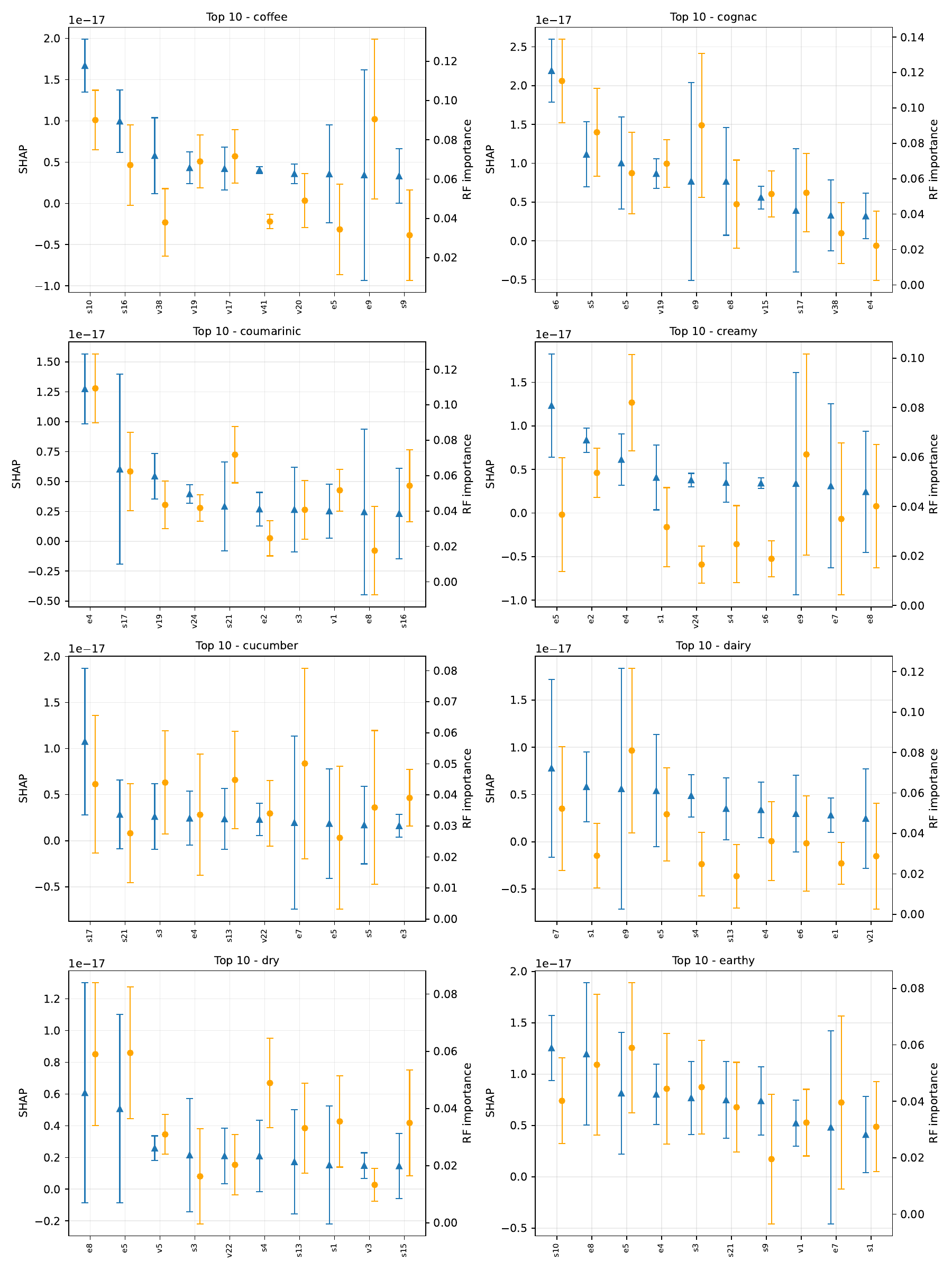}
            \caption{Odors comparison}
\end{figure}

\newpage

\begin{figure}
            \centering
            \includegraphics[width=\linewidth]{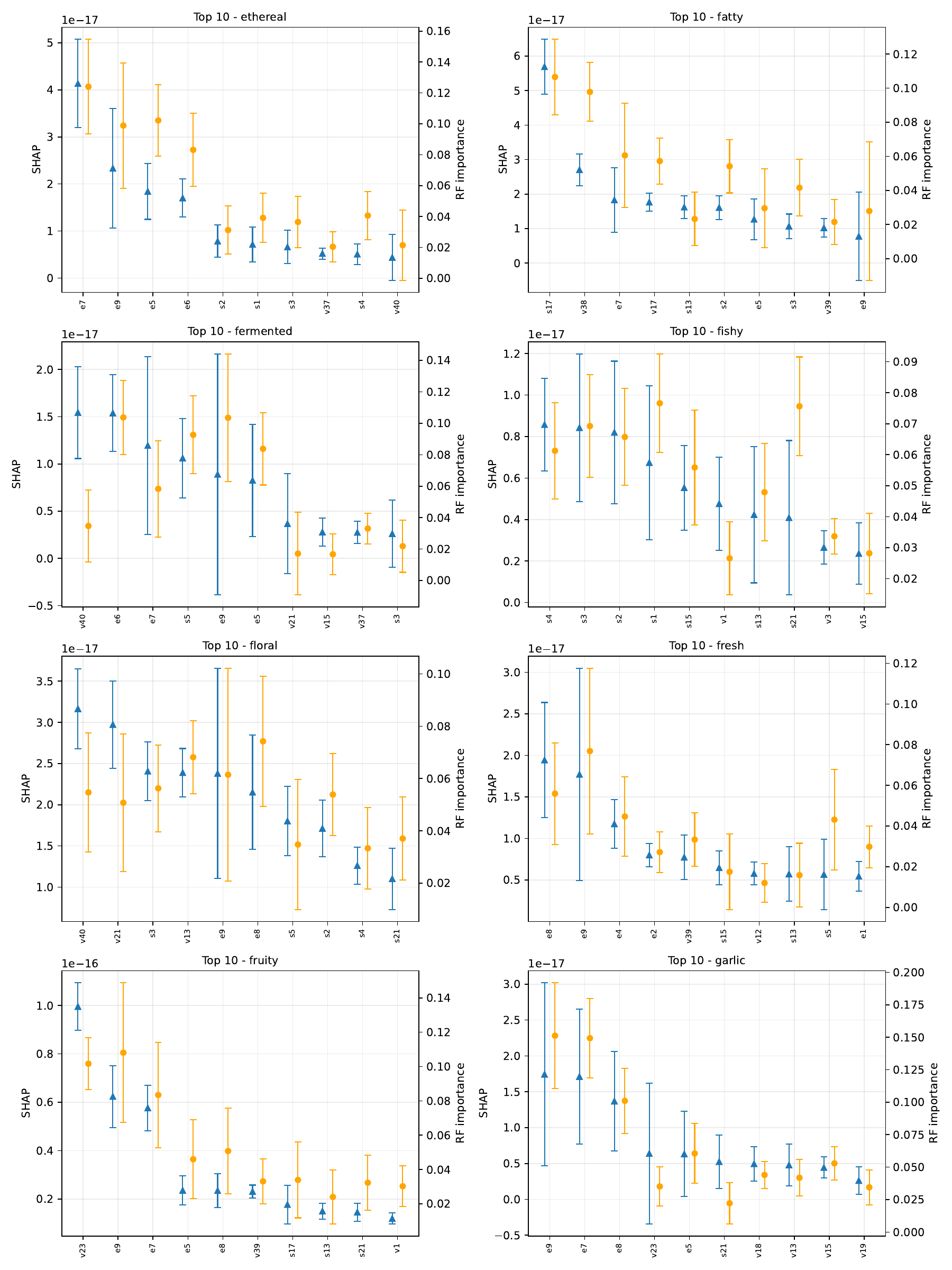}
            \caption{Odors comparison}
\end{figure}

\newpage

\begin{figure}
            \centering
            \includegraphics[width=\linewidth]{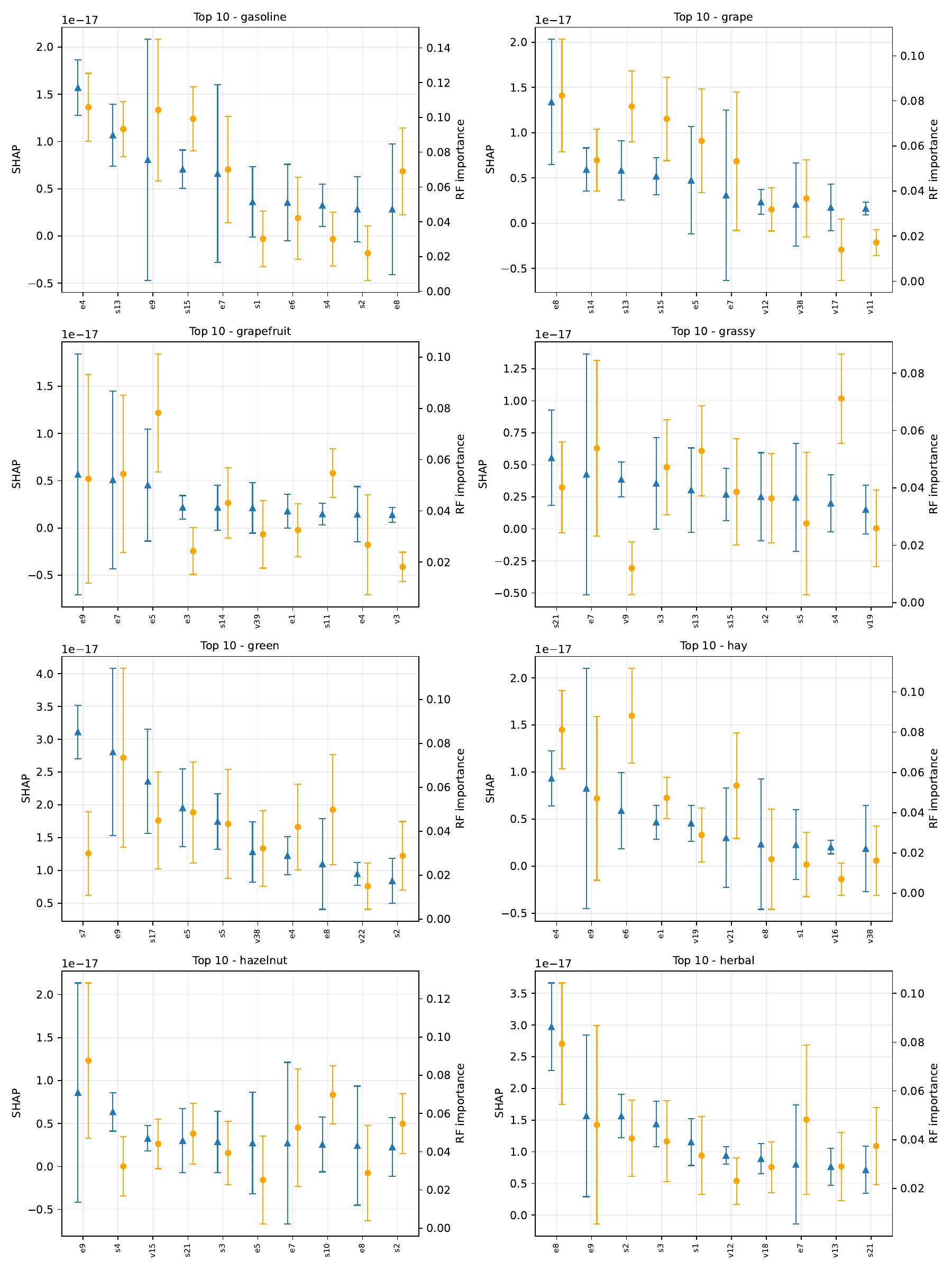}
            \caption{Odors comparison}
\end{figure}

\newpage

\begin{figure}
            \centering
            \includegraphics[width=\linewidth]{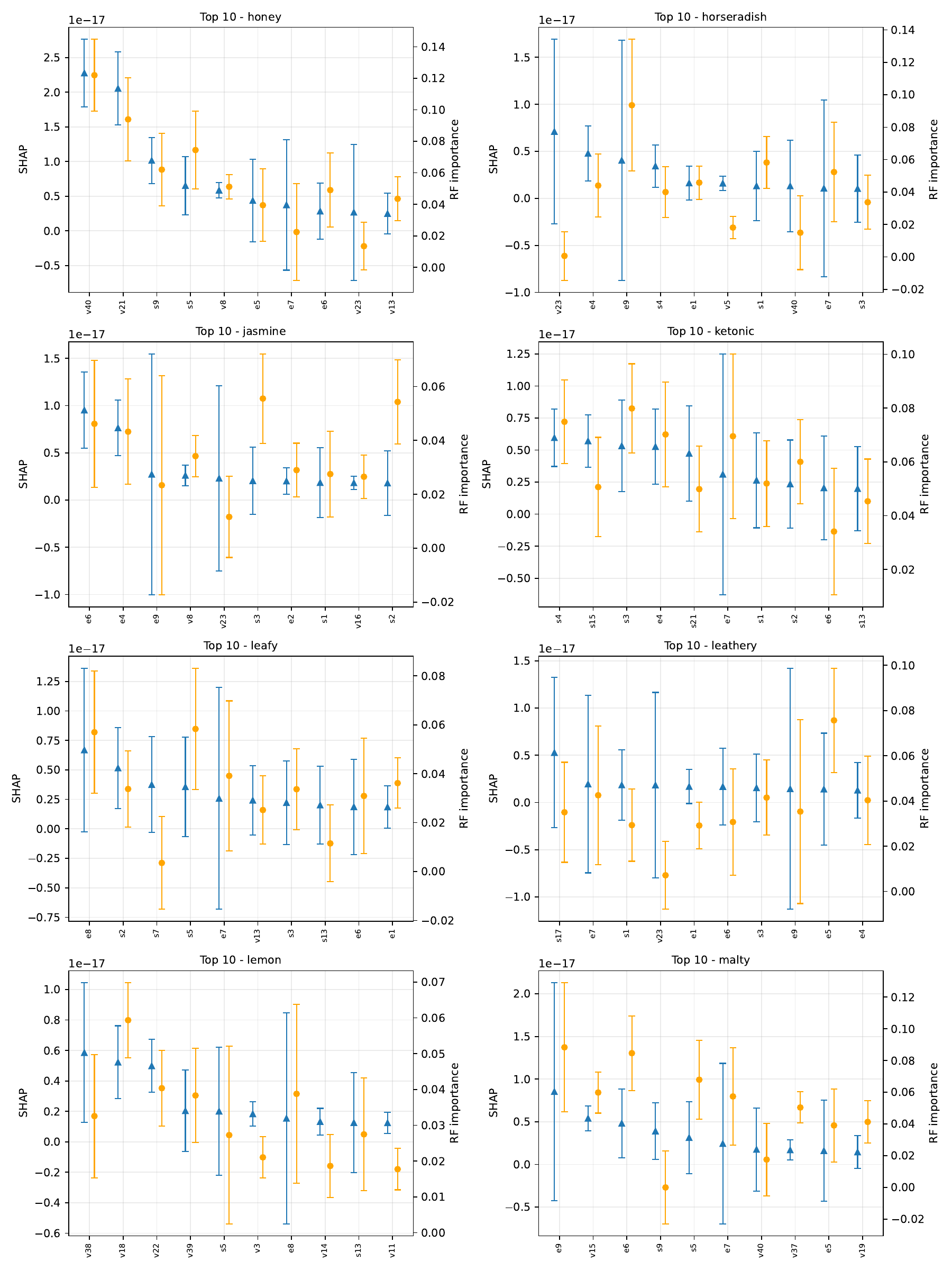}
            \caption{Odors comparison}
\end{figure}

\newpage

\begin{figure}
            \centering
            \includegraphics[width=\linewidth]{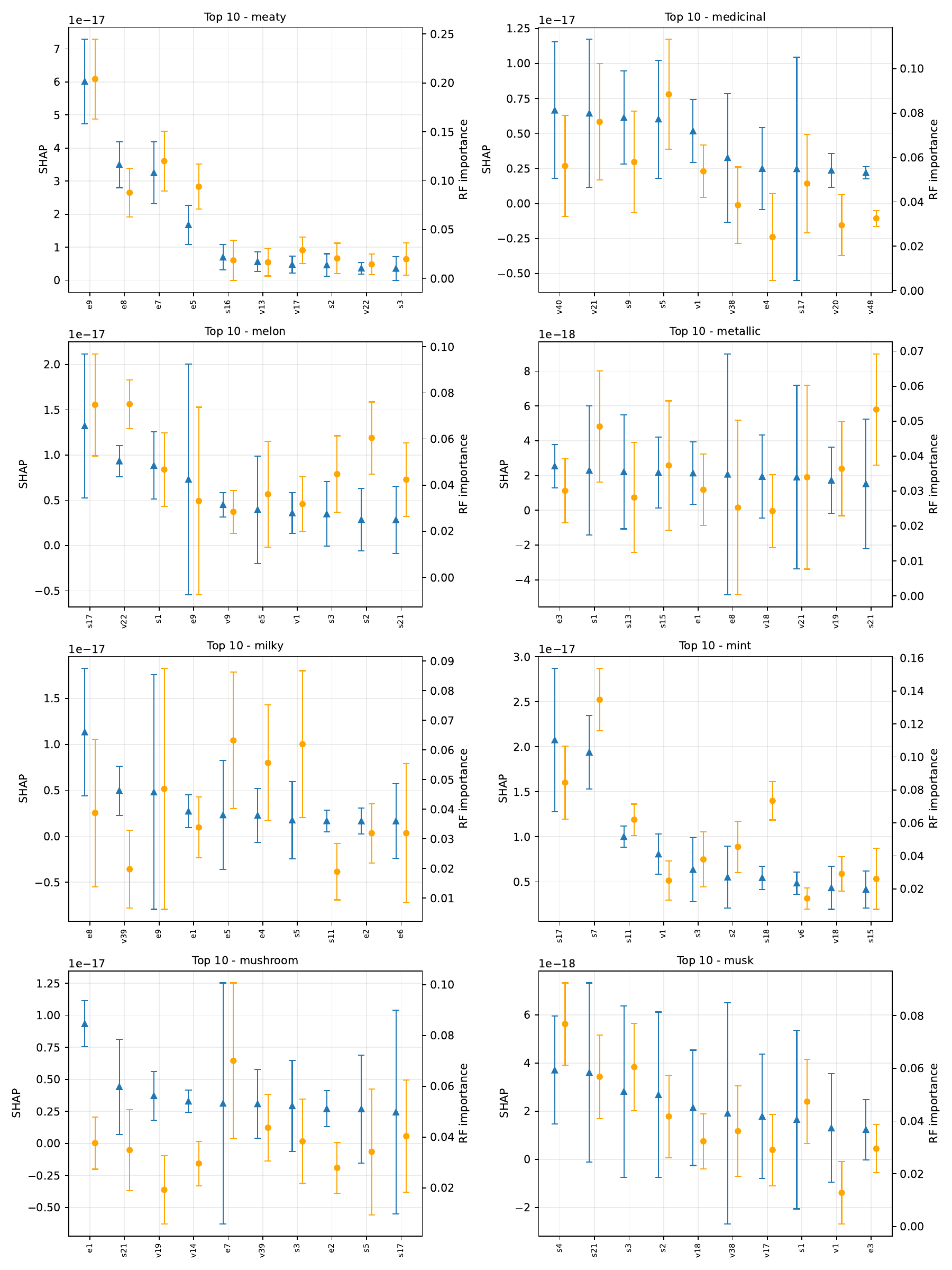}
            \caption{Odors comparison}
\end{figure}

\newpage

\begin{figure}
            \centering
            \includegraphics[width=\linewidth]{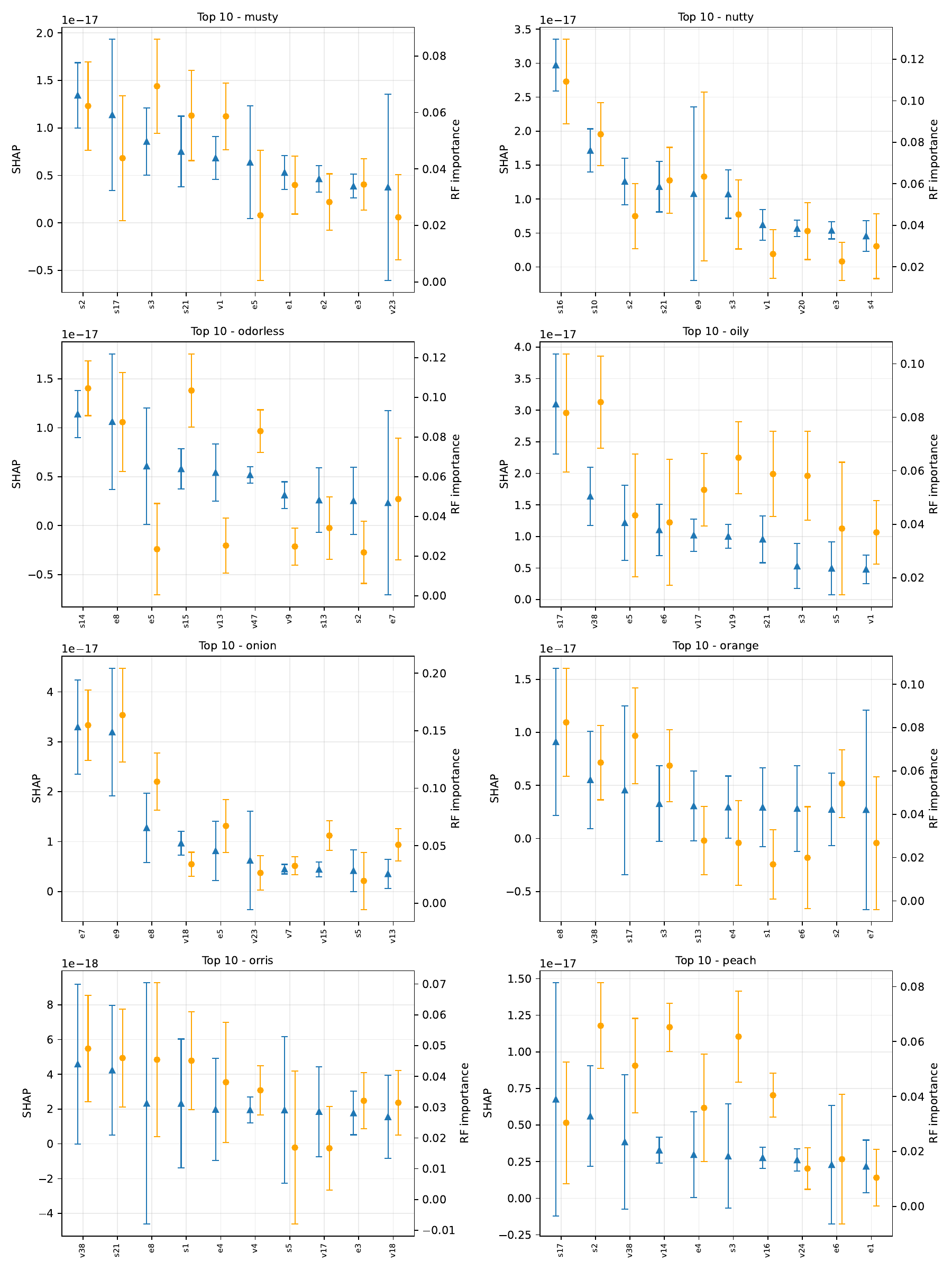}
            \caption{Odors comparison}
\end{figure}

\newpage

\begin{figure}
            \centering
            \includegraphics[width=\linewidth]{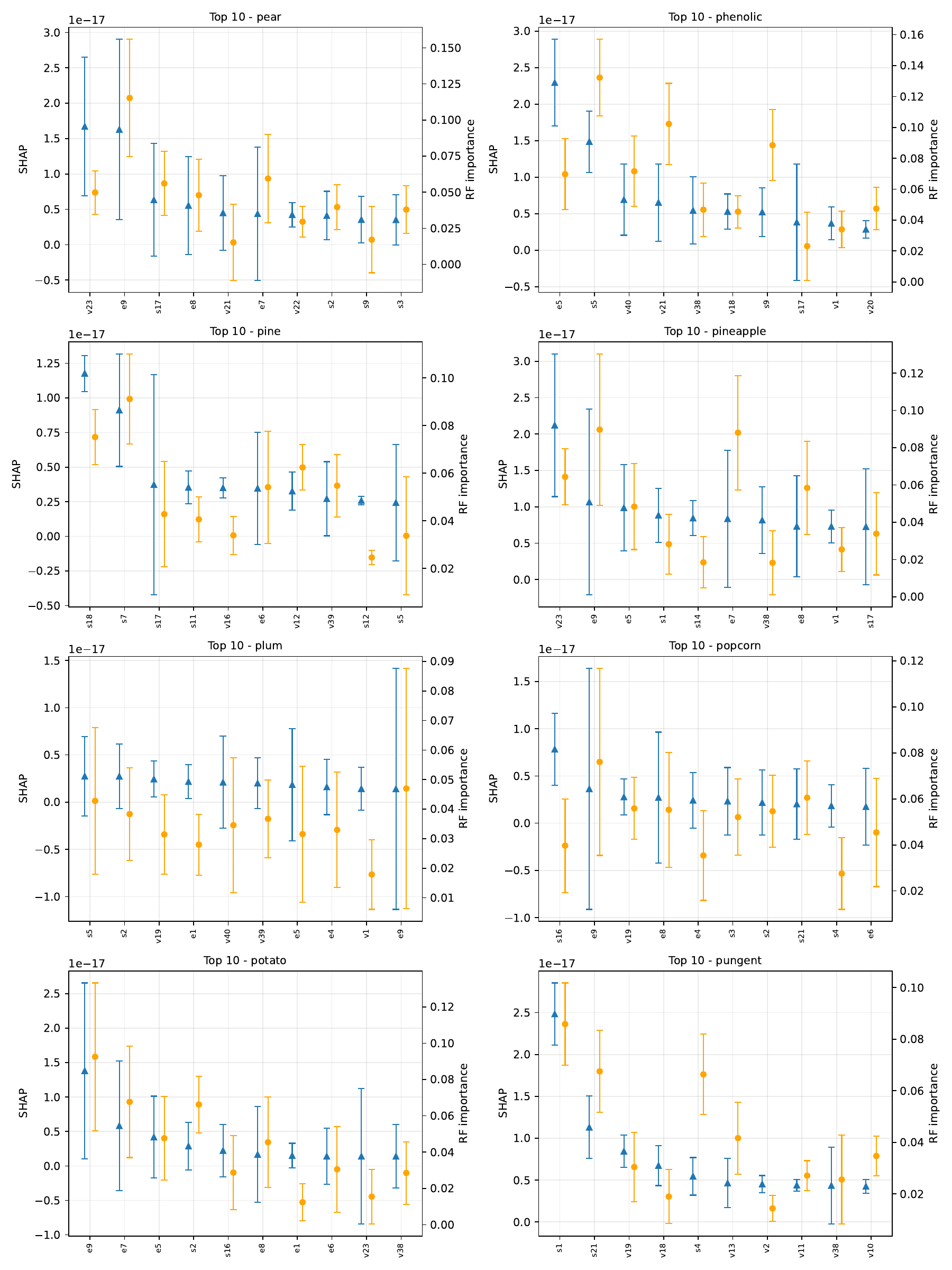}
            \caption{Odors comparison}
\end{figure}

\newpage

\begin{figure}
            \centering
            \includegraphics[width=\linewidth]{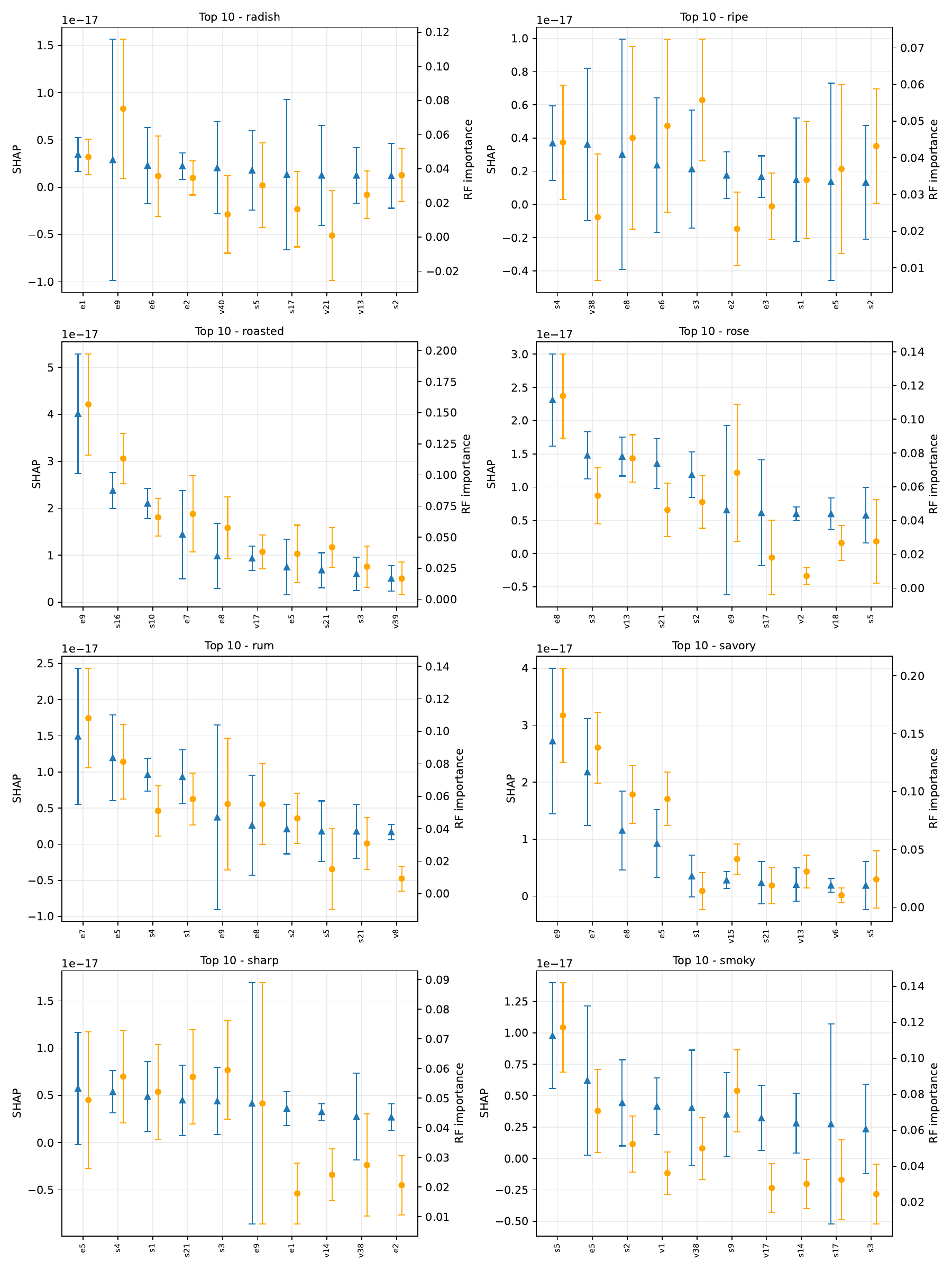}
            \caption{Odors comparison}
\end{figure}

\newpage

\begin{figure}
            \centering
            \includegraphics[width=\linewidth]{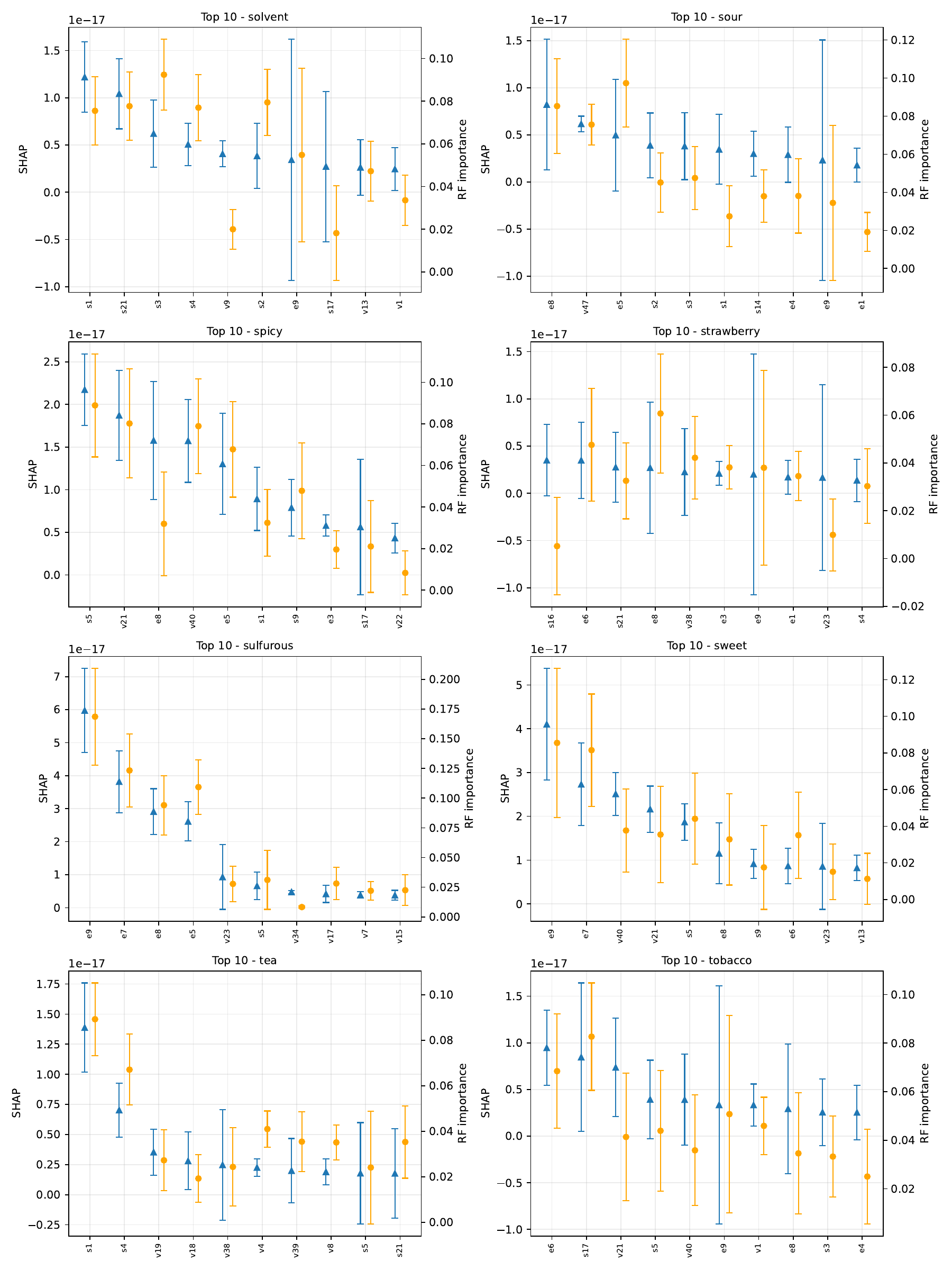}
            \caption{Odors comparison}
\end{figure}

\newpage

\begin{figure}
            \centering
            \includegraphics[width=\linewidth]{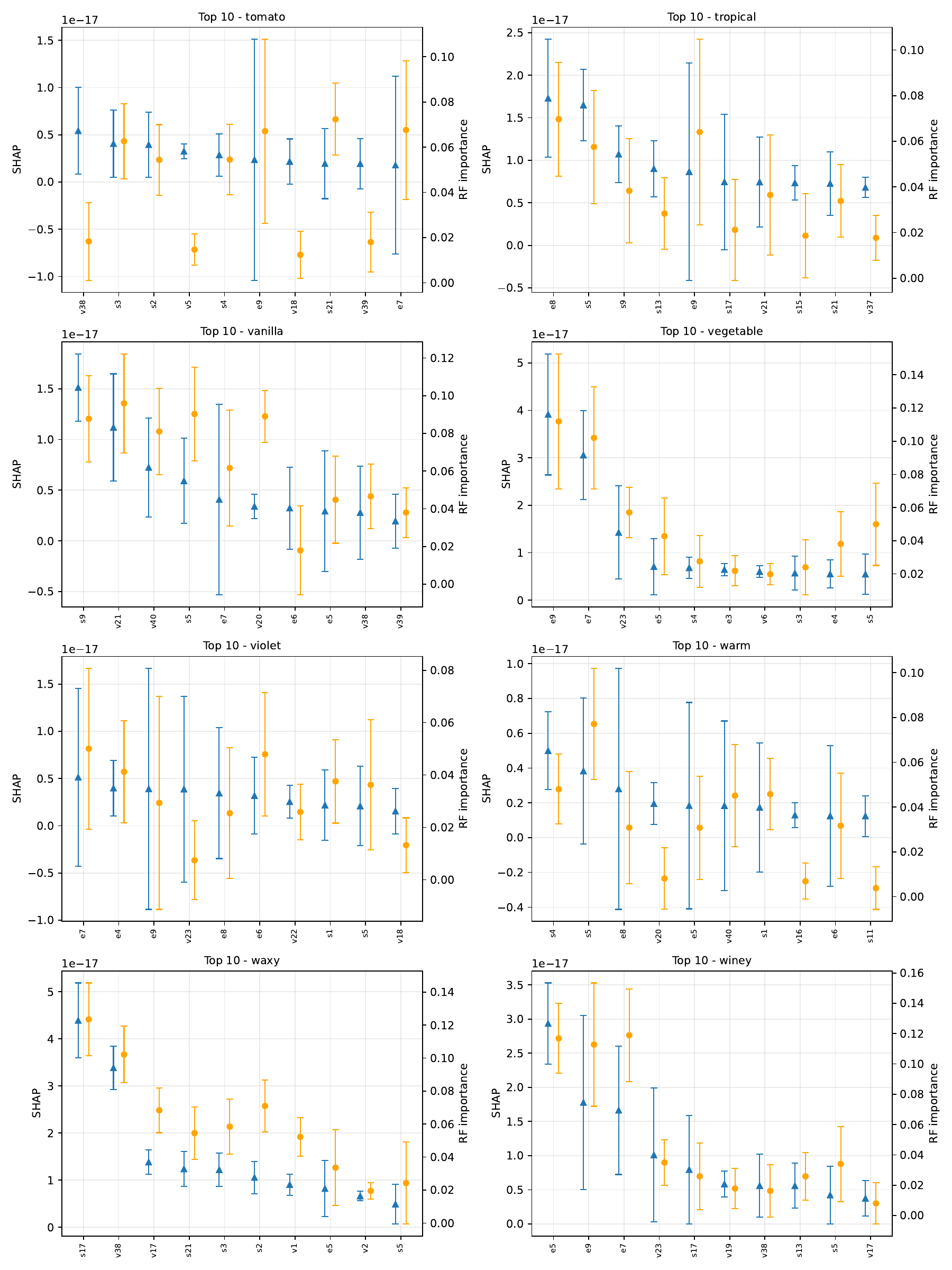}
            \caption{Odors comparison}
\end{figure}

\newpage

\begin{figure}
            \centering
            \includegraphics[width=\linewidth]{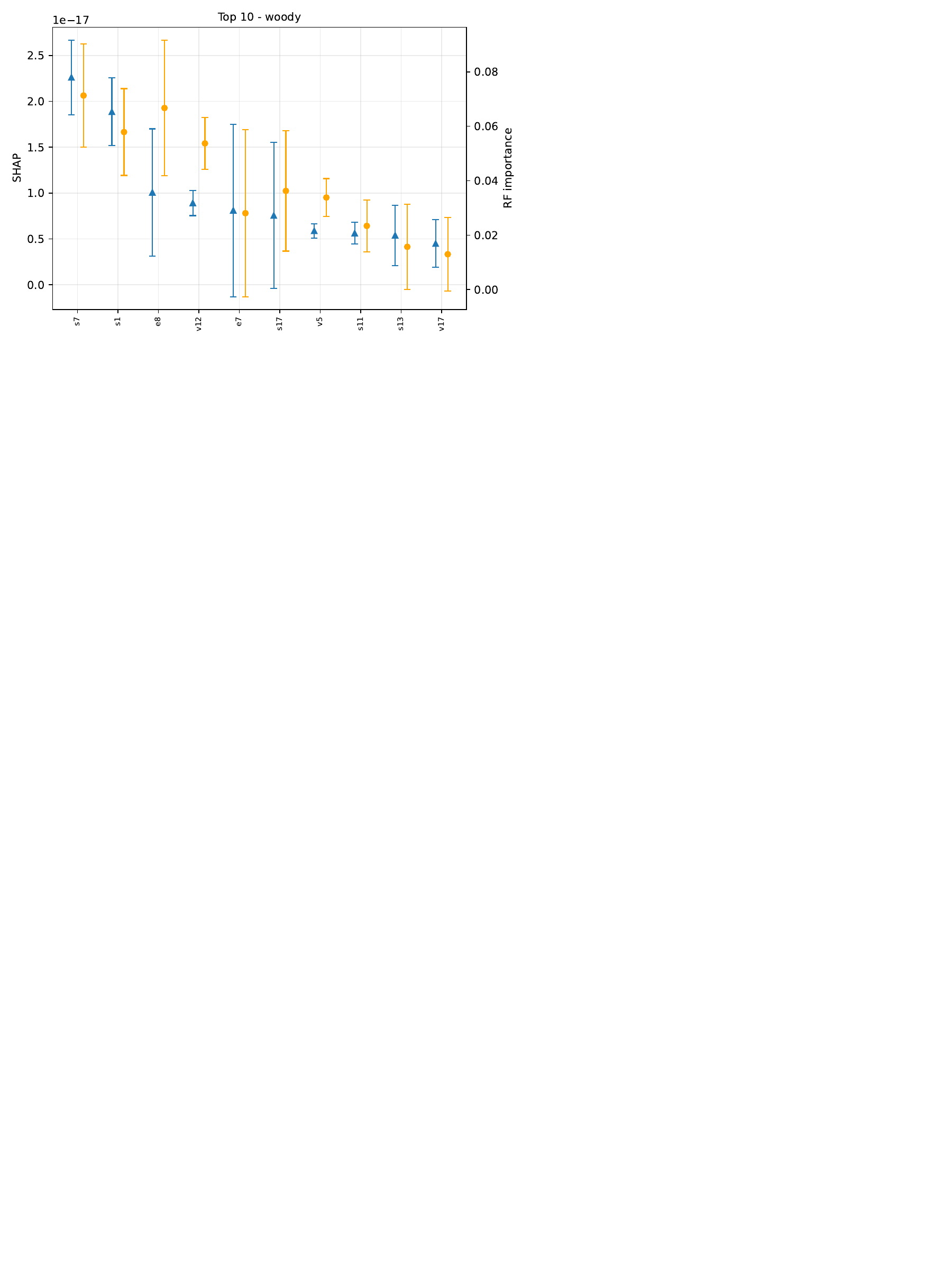}
            \caption{Odors comparison}
\end{figure}

\end{document}